\documentclass[11pt,a4paper,logo]{googledeepmind}

\setleftlogo[110pt]{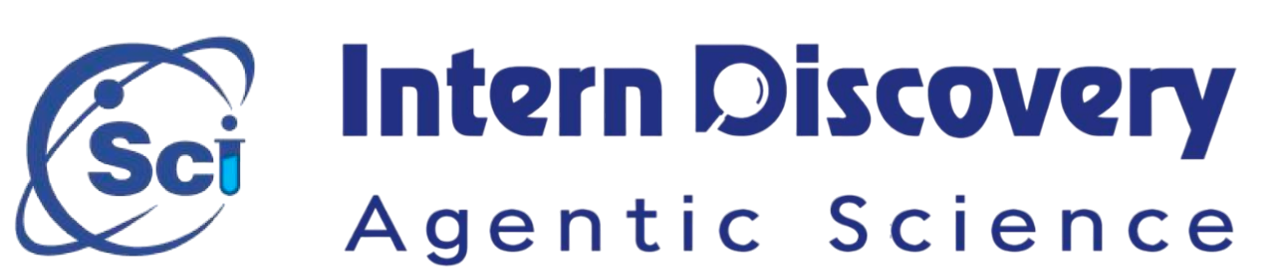}  
\setrightlogo[160pt]{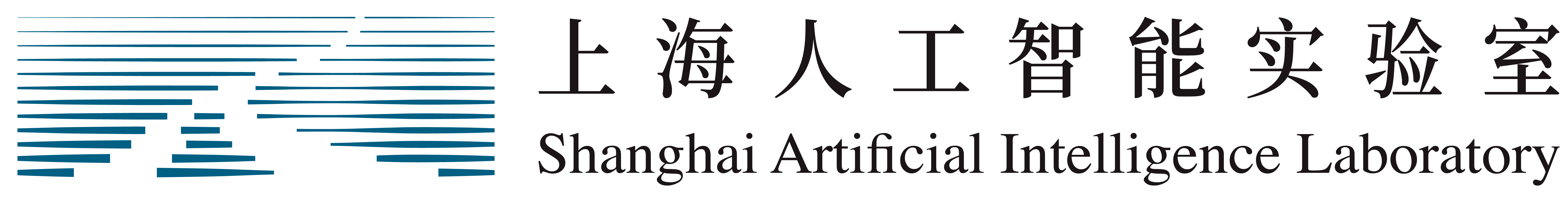}
\usepackage{fontawesome5}
\usepackage{marvosym}
\usepackage{academicons}
\usepackage{etoolbox}
\usepackage[T1]{fontenc}
\usepackage{tgcursor}
\usepackage{pifont}
\usepackage[
    natbib=true,
    backend=biber,
    style=numeric, 
    sorting=none, 
    maxbibnames=5, 
    minbibnames=5  
]{biblatex}
\AtEveryBibitem{\clearfield{month}}
\AtEveryBibitem{\clearfield{day}}
\usepackage{csquotes}

\title{SciDataCopilot: An Agentic Data Preparation Framework for AGI-driven Scientific Discovery}

\newcommand{\homepage}{\raisebox{-1.5pt}{\includegraphics[height=0.9em]{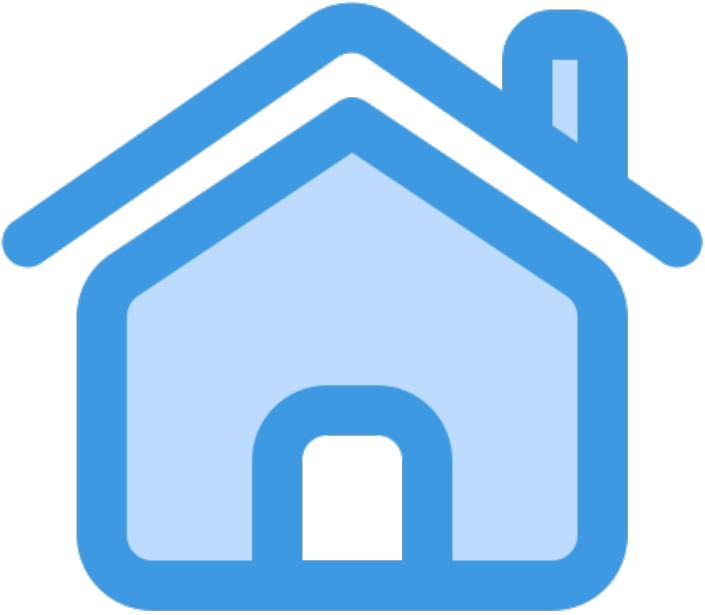}}}

\author{
Jiyong Rao*, 
Yicheng Qiu*, 
Jiahui Zhang*, 
Juntao Deng*, 
Shangquan Sun*,
Fenghua Ling, Hao Chen, \quad Nanqing Dong, Zhangyang Gao, Siqi Sun,
Yuqiang Li, Dongzhan Zhou, Guangyu Wang, Lijun Wu, Conghui He,
Xuhong Wang, Jing Shao, Xiang Liu, Yu Zhu, Mianxin Liu, Qihao Zheng,
Yinghui Zhang, Jiamin Wu, Xiaosong Wang, Shixiang Tang, Wenlong Zhang,
Bo Zhang, Wanli Ouyang,
Runkai Zhao~\Letter, 
Chunfeng Song~\Letter, 
Lei Bai~\Letter, 
Chi Zhang~\Letter~$\dagger$ \\
\centering{\normalsize Shanghai Artificial Intelligence Laboratory}\\
$*$ Co-first Authors~~  \Letter\ Corresponding Authors~~ $\dagger$ Project Lead\\
\faEnvelope[regular]~\texttt{zhangchi2@pjlab.org.cn} \quad
\homepage~\href{https://scidatacopilot.github.io/SciDataCopilot-project/}{\text{ProjectPage}}
}

\usepackage{pdflscape}

\usepackage{textcomp}
\usepackage{rotating}

\usepackage{setspace}
\usepackage{microtype} 

\usepackage{graphicx}
\usepackage{tabularx}
\usepackage{subcaption}
\usepackage{caption}

\usepackage{booktabs}
\usepackage[table]{xcolor}
\usepackage{xcolor}

\usepackage{array}
\usepackage{threeparttable}
\usepackage{multirow}
\usepackage{bm}

\usepackage{amsmath}
\usepackage{siunitx}

\usepackage{enumitem}
\usepackage{float}
\usepackage{seqsplit}
\usepackage{framed}

\usepackage{tikz}

\usepackage{caption} 
\usepackage[export]{adjustbox}

\usepackage{ragged2e}
\usepackage[most]{tcolorbox}

\usepackage{makecell}
\usepackage{tabularray}

\usepackage{longtable}
\usepackage{listings}
\definecolor{CaseGreen}{HTML}{2E7D32} 
\definecolor{CaseOrange}{HTML}{F57C00} 
\definecolor{CaseGray}{HTML}{F6F7F8}   
\definecolor{CaseInk}{HTML}{212121}    
\definecolor{CaseWhite}{HTML}{F5F5DC}
\definecolor{DeepBlue}{HTML}{003366} 
\definecolor{LightBlue}{HTML}{99CCFF} 
\definecolor{DeepPurple}{HTML}{673AB7} 
\definecolor{MiddlePurple}{HTML}{9C7FD0}
\definecolor{LightPurple}{HTML}{D1C4E9} 
\definecolor{HotPink}{HTML}{FF69B4} 
\definecolor{SoftPink}{HTML}{F8BBD0} 
\definecolor{Crimson}{HTML}{DC143C} 
\definecolor{Teal}{HTML}{008080} 
\definecolor{Cyan}{HTML}{00BCD4} 
\definecolor{SoftGray}{HTML}{EEEEEE}       
\definecolor{LighterGray}{HTML}{FAFAFA} 

\newtcolorbox{stagebox}[1]{
  breakable, enhanced,
  colback=CaseGray, colframe=DeepBlue, coltitle=CaseWhite,
  title=\bfseries #1, fonttitle=\bfseries,
  left=1.2mm,right=1.2mm,top=1.2mm,bottom=1.2mm, boxrule=0.6pt
}

\newtcblisting{verbatimbox}{
  listing only, breakable, enhanced,
  colback=white, colframe=LightBlue, boxrule=0.5pt,
  left=1.2mm,right=1.2mm,top=1.2mm,bottom=1.2mm
}

\usepackage[symbol]{footmisc}
\newcolumntype{Y}{>{\RaggedRight\arraybackslash}X}

\usepackage{hyperref}
\hypersetup{
    colorlinks=true,
    linkcolor=blue, 
    citecolor=blue,  
    filecolor=black,
    urlcolor=blue    
}

\usepackage{tocloft}

\usepackage{etoolbox}
\usepackage{arydshln}
\makeatletter
\patchcmd{\@tocline}
    {\hfil}
    {\leaders\hbox{\hfil}\hfil}
    {}{}
\makeatother

\begin{abstract}
\vspace{-0.6cm}

The current landscape of AI for Science (AI4S) is predominantly anchored in large-scale textual corpora, where generative AI systems excel at hypothesis generation, literature search, and multi-modal reasoning. However, a critical bottleneck for accelerating closed-loop scientific discovery remains the utilization of raw experimental data. Characterized by extreme heterogeneity, high specificity, and deep domain expertise requirements, raw data possess neither direct semantic alignment with linguistic representations nor structural homogeneity suitable for a unified embedding space. The disconnect prevents the emerging class of Artificial General Intelligence for Science (AGI4S) from effectively interfacing with the physical reality of experimentation. In this work, we extend the text-centric AI-Ready concept to \textit{Scientific AI-Ready data paradigm}, explicitly formalizing how scientific data is specified, structured, and composed within a computational workflow.
To operationalize this idea, we propose \textit{SciDataCopilot}, an autonomous agentic framework designed to handle data ingestion, scientific intent parsing, and multi-modal integration in a end-to-end manner. By positioning data readiness as a core operational primitive, the framework provides a principled foundation for reusable, transferable systems, enabling the transition toward experiment-driven scientific general intelligence.
Extensive evaluations across three heterogeneous scientific domains show that SciDataCopilot improves efficiency, scalability, and consistency over manual pipelines, with up to 30$\times$ speedup in data preparation.

\end{abstract}

\begin{document}

\sloppy 
\maketitle


\begin{center}
    \centering
    \captionsetup{type=figure}
    \includegraphics[width=0.95\linewidth]{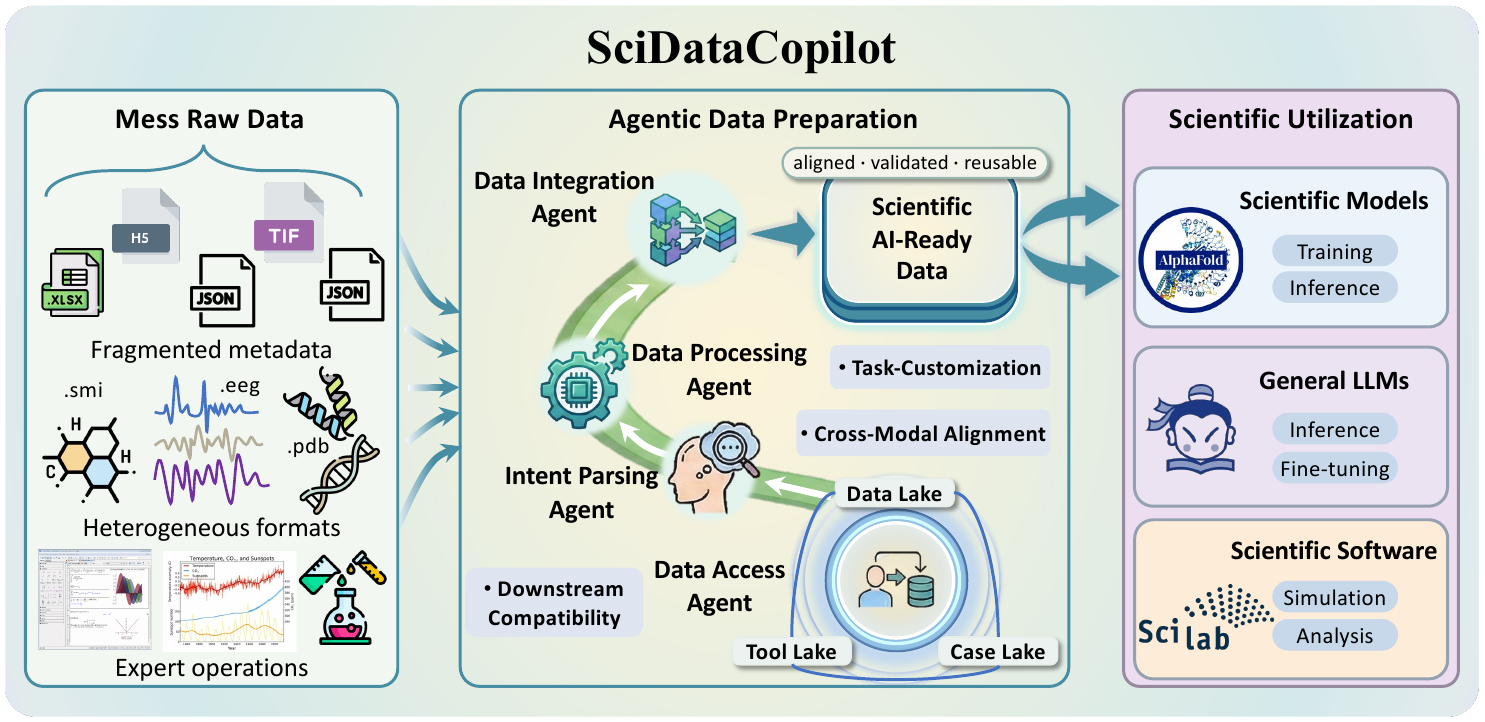}
    \captionof{figure}{The \textbf{Scientific AI-Ready paradigm} formalizes raw data as task-conditioned, cross-modal align for autonomous scientific discovery. The proposed \textbf{SciDataCopilot} instantiates this paradigm through an agentic framework for transforming heterogeneous raw data into Scientific AI-Ready data.}
    \label{fig:teaser}
\end{center}%

\clearpage

\tableofcontents
\enlargethispage{1cm} 
\thispagestyle{fancy} 

\clearpage

\section{Introduction}
\label{sec:intro}
Generative AI~\cite{jaech2024openai,abdin2024phi,Guo2025deepseek,Gemini3_DeepMind2025,yang2025qwen3} is catalyzing a paradigm shift in scientific discovery, evolving traditional research practices into the new era of AI for Science (AI4S).
Its impact permeates multiple stages of the research lifecycle, including hypothesis formulation~\cite{moose2025,team2025internagent}, deep literature search~\cite{du2025deepresearchbenchcomprehensivebenchmark,xu2025probing}, dry/wet experimental design~\cite{watson2023novo,dauparas2022robust}, and evidence interpretation~\cite{zhou2025scientistsexamprobingcognitive,wang2025scireasoner}.
Representative works such as Intern-S1~\cite{bai2025intern} and ScienceOne-AI~\cite{s1agent2025} provide empirical evidence that scientific intelligence emerges at scale when conditioned on comprehensive text-centric archives of papers, patents, and technical reports.
Leveraging such vast scientific corpora as their primitive data source, Multi-Agent Systems, such as InternAgent~\cite{team2025internagent}, AI Co-Scientist~\cite{gottweis2025towards,ai_coscientist_framework} and SciAgents~\cite{ghafarollahi2025sciagents}, are capable of automating end-to-end research workflows. These agents collaboratively orchestrate complex discovery pipelines, executing tasks from ideation to validation bypassing traditional human-centric methods.

However, the conventional AI-Ready paradigm linearizes structured scientific data, forcing inherently complex representations into sequential formats optimized for LLM-centric alignment learning.~\cite{zhou2025scientistsexamprobingcognitive,gridach2025agentic,zhou2025323}
In practice, as illustrated in Figure~\ref{fig:teaser}, scientific data take diverse forms, ranging from structured molecular sequences constrained by chemical and biological validity, to high-dimensional neural recordings governed by protocol-dependent preprocessing, and multi-source observational data requiring explicit spatial-temporal alignment.
This complexity exposes a fundamental gap: a standardized data paradigm is absent to systematically harmonize these distinct modalities into a format directly consumable by downstream scientific analysis\cite{jumper2021highly}.

To bridge this gap, we introduce the \textbf{Scientific AI-Ready data paradigm}, a unified formulation that anchors AI-based scientific modeling and reasoning in first-principles. Unlike text-centric formats, it transforms raw heterogeneous data into structured representations of scientific objectives, domain-specific constraints, and procedural experimental logic.
This, in turn, establishes a necessary foundation for advancing from task-conditioned AI4S toward Artificial General Intelligence for Science (AGI4S), positioning AI as a collaborative partner that synergizes with scientist intuition to accelerate scientific breakthrough.

Compared to conventional AI-Ready data paradigms that prioritize data cleanliness and LLM-centric fitting, the Scientific AI-Ready data paradigm re-defines how scientific data are specified, organized, and consumed to maximize scientific usability. The paradigm is characterized as follows:
\begin{itemize}
    \item \textbf{Task-conditioned principle:} Scientific AI-Ready data paradigm adopts \textbf{scientific tasks as the primary organizing principle}, translating scientific intent to required data units, variables, and constraints. This shifts task-conditioned data specification from inefficient manual collection to automating reusable workflows.
    \item \textbf{Downstream compatibility:} Scientific AI-Ready data paradigm prioritizes direct \textbf{compatibility with downstream scientific analysis}, ensuring that prepared data satisfy model-specific input constraints and enabling composable, executable workflows beyond standalone inference. 
    \item \textbf{Cross-integration ability:} Scientific AI-Ready data paradigm emphasizes \textbf{principled cross-modal and cross-disciplinary alignment}, enabling systematic data association, retrieval, alignment, and composition across heterogeneous scientific domains. 
\end{itemize}
Therefore, the above distinctions reflect a paradigm shift in the underlying objective of scientific data moving from model fitting data toward scientific task-conditioned, constraint-consistent data that can be directly utilized by scientific analysis.

Operationalizing the Scientific AI-Ready data paradigm introduces substantial challenges in both breadth and depth.
First, scientific data preparation is inherently task-conditioned and originates from diverse sources. It causes diverse objectives leading to fragmented data assets and processing logic across experiments and domains.
Besides, scientific data exhibit high structural and semantic complexity, depending on task intent, data properties, and domain knowledge.
The difficulties are further exacerbated by the requirement to align and compose data across modalities, scales, and experimental contexts.
These challenges are amplified when data must be associated, aligned, and composed across heterogeneous modalities, scales, and experimental contexts.
As a result, these characteristics make monolithic workflows insufficient, as they lack the flexibility to adapt processing logic to evolving task requirements and data conditions.
Instead, they motivate a staged yet adaptive workflow in which data access, intent parsing, processing, and integration are collaborated through collaborative agents, each making context-aware decisions while operating in a holistic manner.

Based on this observation, we propose \textbf{SciDataCopilot}, a unified agentic framework that instantiates the Scientific AI-Ready data preparation through a staged workflow. The overview of the framework is illustrated in Figure~\ref{fig:framework}, which comprises four coordinated agents: a \emph{Data Access Agent} that establishes structured data associations, an \emph{Intent Parsing Agent} that retrieves and plans task-relevant data, a \emph{Data Processing Agent} that performs domain-aware transformations under explicit constraints, and a \emph{Data Integration Agent} that aligns and composes processed data units across modalities and experimental contexts. Together, these agents implement an end-to-end pipeline from data association and retrieval to processing and integration, producing datasets that are directly consumable by downstream scientific models and workflows. More detailed descriptions are given in Section~\ref{sec:design}.

Extensive experiments across heterogeneous scientific domains demonstrate the effectiveness and generalization.
In life sciences, SciDataCopilot enables fully autonomous preparation of large-scale enzyme catalysis datasets from a single natural-language instruction, producing 214K Scientific AI-Ready enzyme–reaction records within hours, achieving an almost 20× increase in scale over prior collected databases.
In neuroscience, SciDataCopilot standardizes EEG/MEG analysis across four sub-tasks, including alpha extraction, EOG regression, ICA decomposition, and large-scale EEG preprocessing. It matches human-expert quantitative performance while delivering 3–5$\times$ faster end-to-end execution and structured execution traces.
In earth sciences, SciDataCopilot executes cross-disciplinary meteorological data preparation under strict temporal constraints, achieving over 30× efficiency gains relative to manual spreadsheet-based workflows.
Together, these results highlight SciDataCopilot as a scalable and domain-agnostic framework for transforming heterogeneous raw scientific data into Scientific AI-Ready data, bridging human intent and automated, quality-aware data preparation across disciplines. We are actively validating the framework across a broader range of scientific domains.
Overall, our key contributions are summarized as follows:
\begin{itemize}
    \item \textbf{Task-conditioned Scientific AI-Ready Data Paradigm.} We introduce \emph{Scientific AI-Ready data} as a new data paradigm that generalizes conventional AI-Ready formats. This paradigm explicitly formalizes task-conditioned data specification, scientifically grounded data structuring, and cross-domain data composition, enabling complex scientific data to be systematically aligned with downstream scientific discovery.
    \item \textbf{Unified Agentic Framework for Scientific Data Preparation.} We propose \emph{SciDataCopilot}, a unified agentic framework that decomposes scientific data preparation through data access, intent parsing, data processing, and data integration. The framework enables task-driven data customization across heterogeneous sources.
    \item \textbf{Cross-Domain Generalization on Scientific Data.} We validate the proposed paradigm and framework across diverse scientific domains, including life science, neuroscience, and earth science. These results demonstrate that Scientific AI-Ready data paradigm supports consistent data preparation across domains with fundamentally different data modalities, structures, and experimental constraints.
\end{itemize}
\section{Related Works}
\label{relatedWork}

Data infrastructure in scientific discovery has long been built around a main line of resource aggregation, metadata registration, search and download, and sharing governance, popular works including Intern-Discovery~\cite{interndiscovery},  OpenDataLab~\cite{he2024opendatalab}, ScienceDataBank~\cite{chengzan2017sciencedb}, and SciHorizon~\cite{qin2025scihorizon}. These works provide large-scale catalogs and basic metadata services that address questions such as `\emph{ where the data are} ' and `\emph{ how to access them }'. However, the delivered artifacts are most often either text-centric outputs (e.g., papers, reports, or structured textual summaries) or lightly processed derivatives of conventional multimodal data (audio, images/videos, and text), rather than analysis-ready scientific datasets. Consequently, broader scientific data modalities are often missing, including time-series measurements, geospatial and remote-sensing rasters, and molecular/structural representations.Addressing a concrete research question still requires substantial downstream work, including cleaning, structuring, alignment, annotation, and AI-Ready input.
Moreover, both general-purpose and domain-specific data preparation approaches still suffer from limited adaptability and portability, as well as high interoperability costs, making it difficult to support scalable processing and consistent integration for cross-disciplinary, multi-modal scientific data.

\subsection{General-Purpose Automated Data Preparation Frameworks} 

Data processing in the general domain has evolved rapidly, with a representative direction being operatorized and pipeline-based data production systems. DataFlow~\cite{liang2025dataflow} places LLMs at the center of the data processing ecosystem and provides reusable operators and general-purpose pipelines for cleaning, filtering, formatting, and synthesizing high-quality training data. Systems such as Data-Juicer~\cite{djv1, chen2025datajuicer, chen2025datajuicer2.0} , Curator~\cite{curator}, and DataTrove~\cite{penedo2024datatrove} also offer rich capabilities for data profiling, deduplication, filtering, format conversion, and multi-modal processing, with strong emphasis on engineering scalability and extensible execution backends. Open source data preparation frameworks such as  Deequ~\cite{schelter2018automating} provide data quality checks, profiling, and validation. Cloud native and big data computation infrastructures such as Apache Spark~\cite{spark2018apache} and Apache Flink~\cite{carbone2015apache} provide scalable execution engines for large-scale data preparation, supporting both batch and streaming scenarios, but additional work is still needed for semantic understanding of scientific data, experimental context, and integration with domain tool chains. 

The systems significantly reduce the barrier to producing general corpora and multi-modal training data, and they promote an industrial approach based on data recipes and reusable pipelines. However, when applied to scientific data, these frameworks often face several difficulties. Scientific data is frequently deposited in file systems as highly heterogeneous experimental packages rather than flat sample sets. The processing chain depends on domain software and models, which generic operators cannot fully cover. Scientific data preparation is not only for training set construction, but also for delivering datasets customized to research intent, supporting reproducible experimental processes, and enabling cross-modal and cross-disciplinary alignment and integration.

Therefore, beyond general data preparation paradigms, scientific discovery requires structured metadata understanding, task-driven planning, and principled links to domain tools and models.

\subsection{Domain-specific Automated Data Preparation Frameworks}

In domain-specific communities, disciplines such as bioinformatics (e.g., the AlphaFold ecosystem)~\cite{AlphaFold-Multimer2021}, earth science (e.g., eo4ai, earthlink)~\cite{eo4ai2022,guo2025earthlink}, and astrophysics have established mature pipelines for processing massive multi-modal data. Domain toolkits such as NanoPyx~\cite{saraiva2025efficiently}, Astropy~\cite{robitaille2013astropy}, and MDAnalysis~\cite{michaud2011mdanalysis} provide mature parsing, processing, and analysis functionality and have built stable user communities. In recent years, higher-level intelligent systems and industry-grade data processing platforms have also emerged. For instance, EarthLink~\cite{guo2025earthlink} is built upon the Intern-S1 foundation model. It systematically integrates climate-domain data, knowledge, tools, and general-purpose large-model capabilities. Through human-in-the-loop intelligent planning, a self-evolving experimentation platform, and multi-scenario scientific analysis, it lowers the barrier to programming and software usage and enables research workflows driven by natural-language instructions. In large-scale biomedical cohort settings, in-house data processing tools associated with the UK Biobank have developed scalable capabilities across sample and phenotype management, genomic data cleaning and quality control, feature construction, and batch processing~\cite{hanscombe_ukbtools_github}~\cite{mbatchou2021computationally}~\cite{baiwenjia_ukbb_cardiac_github}. Moreover, many disciplines continue to rely heavily on specialized toolchains. Representative examples include MNE-Python~\cite{Gramfort2013MNEPython} for EEG and MEG data in neuroscience,  xarray~\cite{Hoyer2017xarray} and NCO~\cite{NCO_website} for gridded and reanalysis data in Earth and climate science, RDKit~\cite{Landrum2013RDKit} for molecular and reaction data in cheminformatics, and Fiji~\cite{Schindelin2012Fiji} and ImageJ~\cite{Collins2007ImageJ} for data in bioimage analysis.

Despite their relative maturity, these tools and platforms are typically used via domain knowledge-base access and script-based pipelines. They commonly face limitations such as complex and tightly coupled toolchains, poor interoperability across disciplines, and a strong dependence on expert experience. Consequently, practical research workflows still encounter persistent challenges. First, there is a lack of machine-computable understanding of highly heterogeneous file contents, making it difficult to formalize processing steps into standardized procedures that are reproducible and auditable. Second, when research objectives, data sources, or tool versions change, existing workflows are difficult to rapidly migrate and reuse, leading to continued reliance on experts for manual debugging and experience-driven decision making.

\subsection{Research Frontier in Automated Data Preparation}

To bridge the gap between data and knowledge, full-stack agent-based solutions have emerged as a growing trend. For example, Intern-S1~\cite{bai2025intern} and SciReasoner~\cite{wang2025scireasoner} strengthen foundational scientific capabilities through multi-modal training and improved reasoning. Building on these foundations, Intern-Agent~\cite{team2025internagent} constructs a closed-loop multi-agent workflow for scientific research that covers ideation, experimentation, and reflection, enabling autonomous planning and iterative optimization of data processing and experimental pipelines.Related efforts toward more autonomous, end-to-end scientific discovery have also been explored in Kosmos~\cite{mitchener2025kosmos} and AI-Scientist~\cite{lu2024ai}. In parallel, agent paradigms for data manipulation and analysis are rapidly evolving. Data-Copilot~\cite{zhang2023data} decomposes high-level natural language intents into executable toolchain actions, while ChatDB~\cite{hu2023chatdb} focuses on database interactions and maintains an operational \emph{memory stream } to preserve multi-turn consistency and support self-correction, thereby lowering the barrier to complex querying and management. Moreover, general-purpose agent frameworks such as AutoGen~\cite{wu2024autogen} ,LangChain~\cite{mavroudis2024langchain},and LlamaIndex~\cite{Liu_LlamaIndex_2022} facilitate multi-agent collaboration, tool invocation, and workflow orchestration, making them well suited for building cross-source analytics, ETL, and report-generation pipelines. Finally, scientific agent evaluation suites such as SGI-Bench~\cite{xu2025probing} and SciEvalKit~\cite{wang2025scievalkit} provide systematic benchmarks to assess alignment and reliability in realistic research settings. Taken together, this layered stack, from foundation models to agents and evaluation, suggests a broader shift in scientific data processing from assistive tools toward increasingly autonomous discovery. Nevertheless, existing data agents remain largely oriented toward generic tool use, and they often lack systematic semantic grounding for scientific data, reusable and assetized workflow components, and versioned closed-loop pipelines with strong traceability. They also provide limited support for constraint-driven alignment when integrating multi-modal and interdisciplinary data.

To mitigate these issues, case-enhanced workflow generation and agentic closed-loop programming offer a more transferable path. By retrieving previously successful solutions, adapting them to new requirements, and incorporating execution feedback for automatic code repair, these approaches reduce hallucination and improve runnability for complex tasks. In practice, such systems are typically instantiated as a closed-loop pipeline of plan, verify, generate, execute, gather feedback, and iterate. Our work is aligned with this line of thinking, but further systematizes it into a unified architecture tailored for scientific data, as summarized in Table~\ref{tab:system_comparison}.
\begin{itemize}[left=0pt]
    \item \textbf{Data Awareness:} Data access with computable scientific metadata and ontologies, enabling the system to ingest heterogeneous datasets and automatically derive machine-actionable interpretations and dataset descriptions.
    \item \textbf{Case Reuse:} A Case Lake that stores reusable workflow assets. Each case captures data units, toolchains, and integration strategies, supporting retrieval and adaptation to reduce planning uncertainty.
    \item \textbf{Requirements Loop:} A multi-agent closed-loop workflow that translates user requirements into executable pipelines and versioned artifacts, with provenance tracking, failure traceability, and iterative repair.
    \item \textbf{Data Integration:} Constraint-driven integration for cross-modal and cross-disciplinary alignment, where explicit constraints guide harmonization and fusion to generate high-value datasets tailored to the user’s scientific objectives.
\end{itemize}

Overall, this work further extends the paradigm of operator- and pipeline-driven data production to scientific settings. We develop a systematic implementation that integrates a DataFlow-for-Science layer with multi-agent orchestration and an ontology- and case-based knowledge repository, thereby advancing scientific data platforms from mere resource provisioning toward the production of Scientific AI-Ready datasets.

\newcommand{\ccmark}{\textcolor{green!50!black}{$\checkmark$}} 
\newcommand{\ccmarkbf}{\textcolor{green!50!black}{\bm{$\checkmark$}}}
\newcommand{\xxmark}{\textcolor{red!75!black}{$\times$}}   
\newcommand{\ttmark}{\textcolor{blue!70!black}{{$\triangle$}}}
\newcommand{\icon}[1]{%
  \makebox[1.8em][c]{\raisebox{-0.2em}{\includegraphics[height=1em]{#1}}}%
  \hspace{0.4em}%
}
\newcolumntype{Y}{>{\centering\arraybackslash}X}
\renewcommand\theadfont{\bfseries}
\setcellgapes{2pt}\makegapedcells

\begin{table}[t]
\centering
\small
\setlength{\tabcolsep}{4pt}
\renewcommand{\arraystretch}{1.2}
\begin{tabularx}{\linewidth}{l c *{5}{Y}}
\hline
\multicolumn{1}{c}{\thead{System}} &
\thead{Multi-\\domain} &
\thead{Multi-\\modal} &
\thead{Data\\Semantic} &
\thead{Case\\Reusable} &
\thead{Closed\\Loop} &
\thead{Data\\Fusion} \\
\hline

\icon{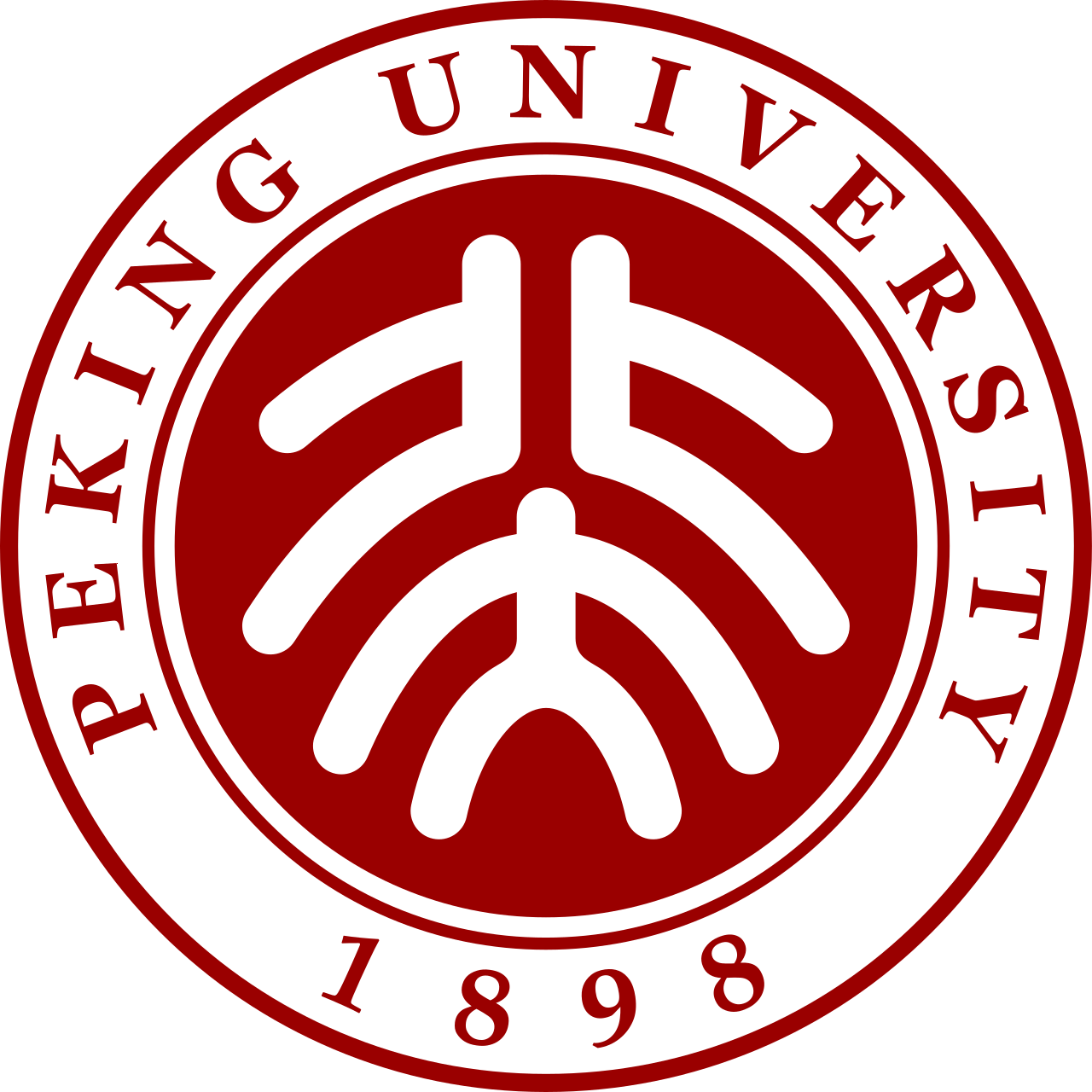} DataFlow~\cite{liang2025dataflow}
& \ccmark & \xxmark & \xxmark & \ttmark & \ttmark & \xxmark \\

\icon{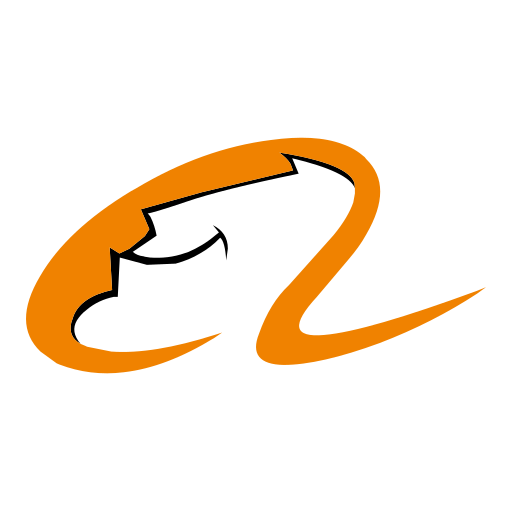} Data-Juicer~\cite{djv1, chen2025datajuicer, chen2025datajuicer2.0}
& \ccmark & \ccmark & \xxmark & \xxmark & \xxmark & \xxmark \\

\icon{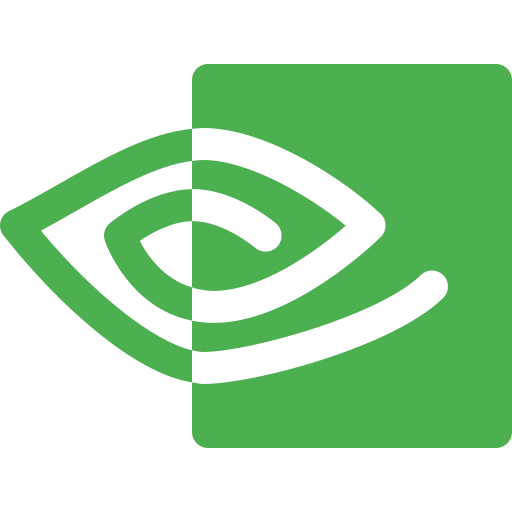} Curator~\cite{curator}
& \ccmark & \ccmark & \xxmark & \xxmark & \xxmark & \xxmark \\

\icon{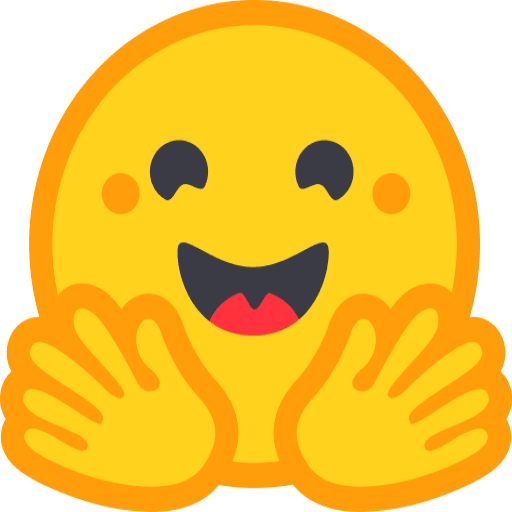} DataTrove~\cite{penedo2024datatrove}
& \ccmark & \xxmark & \xxmark & \xxmark & \xxmark & \xxmark \\

\icon{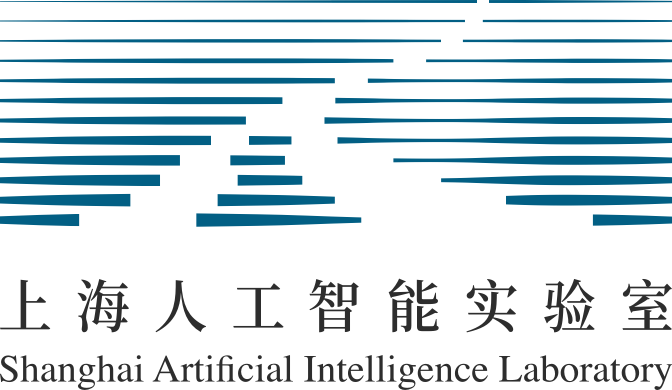} \textbf{SciDataCopilot}
& \ccmarkbf & \ccmarkbf & \ccmarkbf & \ccmarkbf & \ccmarkbf & \ccmarkbf \\
\hline
\end{tabularx}

\caption{Capability comparison between representative general-purpose data preparation frameworks and our SciDataCopilot. Columns summarize whether a system supports cross-domain and multimodal data preparation, scientific data semantics (machine-computable metadata and ontology grounding), case-level workflow reusability, agentic closed-loop iteration, and constraint-driven data fusion for cross-modal integration. \ccmark: supported, \ttmark: partially supported, \xxmark: not supported.}
\label{tab:system_comparison}
\end{table}
\section{Rethinking Data Agents for Scientific Discovery}
\label{sec:rethink}
This section rethinks \emph{Scientific Data Agents} as the substrate that connects scientific questions, diverse data assets, domain operators, and downstream scientific models within end-to-end workflows. We argue that the central bottleneck is not merely data volume or storage, but the lack of a systematic interface for making experimental data scientifically valid, constraint-consistent, and workflow-composable across modalities and disciplines. Consequently, this section systematically elaborates on the core requirements for Scientific Data Agents, analyzes the functional deficiencies of existing data preparation systems, and derives an evolutionary framework to support Agentic Science~\cite{wei2025ai}.

\subsection{Conceptual Formulation}
\indent\textbf{\emph{What is Scientific Data Agent?}}
Scientific Data Agents reflect a shift in AI system design, from models that perform isolated prediction to systems that support autonomous reasoning and execution over scientific data and workflows. These agents, typically built upon scientific large language models~\cite{hu2025survey}, are designed to plan, execute, and iterate across the discovery lifecycle to fulfill high-level research objectives.
Unlike traditional AI assistants, scientific agents autonomously decompose complex goals, such as identifying drug candidates, into structured and executable sub-tasks.
They gather and integrate multi-modal cues, conduct experiments, and synthesize results into coherent scientific discovery.

Crucially, Scientific Data Agents operate over \emph{AI-Ready Data}, which in current AI4S systems typically refer to data that are cleaned, standardized, and formatted for direct consumption by AI models, particularly LLMs. Such data emphasize syntactic consistency and schema normalization, supporting large-scale model fitting and literature-grounded reasoning. However, AI-Ready Data does not explicitly encode scientific task intent, experimental context, or domain constraints, and are therefore insufficient for supporting autonomous, end-to-end scientific workflows.
By adhering to structured, hypothesis-driven workflows that mirror the scientific method, these agents incorporate domain-specific constraints and verification protocols to ensure that autonomous discoveries remain reproducible and scientifically rigorous.

\subsection{Design Implications}
\indent\textbf{\emph{Where do Scientific Data Agents fall short?}}
Current scientific data agents exhibit significant operational and cognitive bottlenecks, primarily characterized by a heavy reliance on textual modalities at the expense of primary empirical data and the absence of failed results in training corpora. These systems demonstrate persistent fragility in long-horizon research tasks, frequently suffering from knowledge expiration due to the static nature of their underlying models and a lack of integrated metadata necessary for assessing data quality and experimental context. Consequently, these agents often fail to achieve high success rates in the orchestration of complex tools or maintain scientific rigor when faced with the rapidly evolving frontier of research discovery.

\indent\textbf{\emph{What Architectural Structure is required?}}
The realization of robust agentic science necessitates a hierarchical architectural structure centered on a three-stage evolutionary framework: establishing a multimodal data foundation, fostering emergent scientific knowledge, and enabling autonomous discovery frontiers. This architecture must incorporate an operating system-level interaction protocol, such as the Model Context Protocol~\cite{mcp} and the Science Context Protocol~\cite{jiang2025scp}, to standardize tools discovery, authorization, and execution across heterogeneous environments like computational simulators and robotic laboratories. Furthermore, the structure requires automated continuous integration pipelines for real-time data ingestion and immutable traceability chains to ensure the transparency, auditability, and reproducibility of the entire autonomous research lifecycle.

Taken together, the above requirements imply a staged architecture for scientific data preparation, which we instantiate as SciDataCopilot in Section~\ref{sec:design}.

\section{SciDataCopilot Framework}
\label{sec:design}
\begin{figure}[t]
    \centering
    \includegraphics[width=1.0\linewidth]{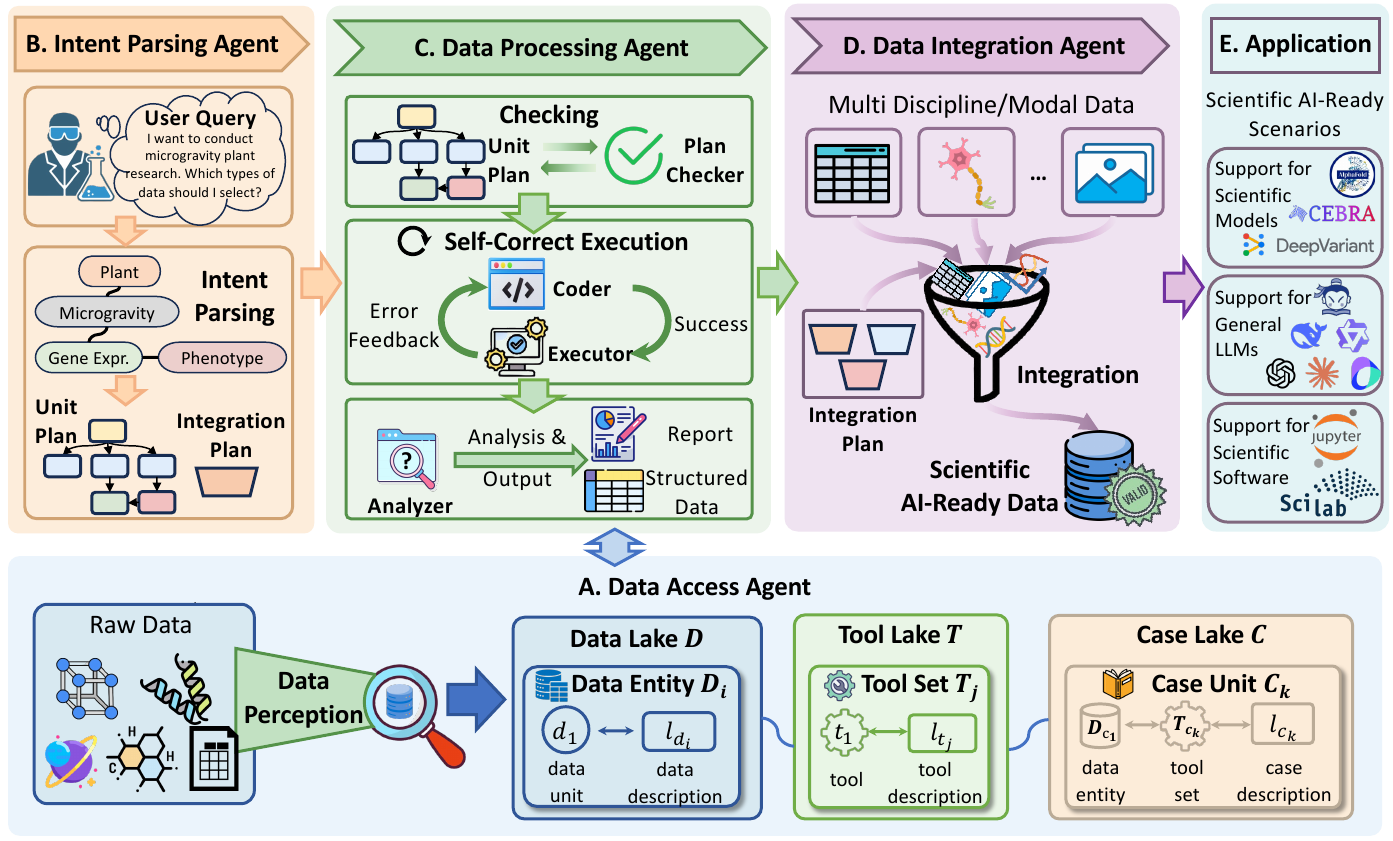}
    \caption{\textbf{Architecture of SciDataCopilot.} The framework integrates four collaborative agents (Data Access, Intent Parsing, Data Processing, and Data Integration) to autonomously align user intents with complex data resources. Ultimately, SciDataCopilot bridges heterogeneous scientific data with specific models, re-defining task-guided data customization and cross-disciplinary integration to empower diverse scientific research tasks.}
    \label{fig:framework}
\end{figure}

SciDataCopilot is a unified, staged multi-agent framework that operationalizes the Scientific AI-Ready paradigm by mapping scientific intent to a composable data-preparation workflow spanning data probing, planning, transformation, and integration. Given a natural language scientific query $q$, the framework operates over a structured knowledge base $\{D, T, C\}$. Here, $D$ represents the re-organized scientific datasets as well as their descriptions, $T$ signifies the available processing tools and specifications, and $C$ captures historical cases connecting scientific questions, data entities, and processing pipelines, including both prepared cases and retrieved cases from external databases.

The overall framework proceeds in four stages as shown in Figure~\ref{fig:framework}. Initially, SciDataCopilot performs data access and knowledge construction by organizing heterogeneous raw data, tools, and historical precedents into a unified knowledge base. Subsequently, during intent parsing and planning, the system interprets the query $q$ to isolate relevant data assets and formulate task-specific processing strategies. Guided by these plans, SciDataCopilot executes data processing pipelines that transform selected data units into task-aligned outputs through domain-aware operations and iterative debugging. Finally, the processed data outputs are integrated according to explicit integration strategies, yielding a coherent and Scientific AI-Ready dataset that satisfies the original scientific intent. The detailed design and functionality of each agent are presented in the following sections.


\subsection{Data Access Agent}
\label{subsec:access}
Scientific data rarely arrive as a unified, standardized table. 
In practice, datasets are accumulated as multi-source, multi-modal, multi-version artifacts distributed across local file systems, shared storage, or exported experiment bundles. 
Even within the same task domain, data produced by different teams or projects often exhibit inconsistent naming conventions, arbitrary directory hierarchies, incompatible field definitions, and highly variable patterns of missing values and outliers, making one-off scripts brittle and hindering scalable automation. 
To address this, we conceptualize data perception not as a simple file-reading routine, but as a collaborative system for path exploration, schema inference, and metadata normalization. 
Building on this view, we develop a \textbf{Data Access Agent} that can autonomously explore data directories and process datasets in a self-organizing manner, continuously extracting and normalizing metadata into a form consumable by downstream automatic planning modules, forming an initial prototype of a scientific data ontology.


\subsubsection{Scientific Database Construction}

\begin{figure}[!t]
    \centering
    \includegraphics[width=1.0\textwidth]{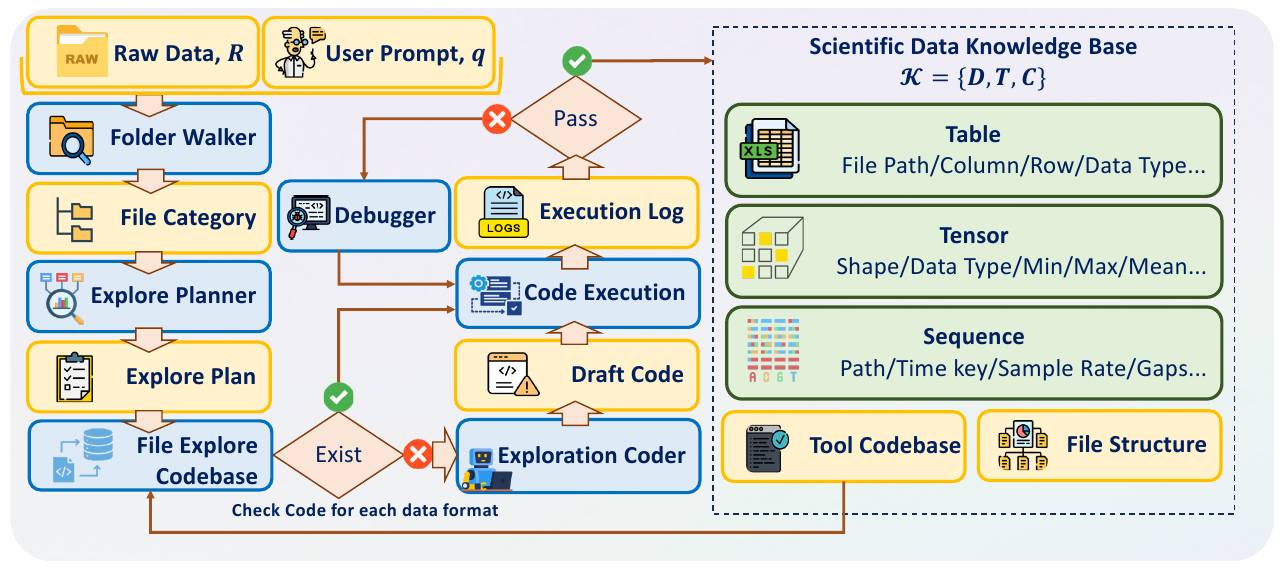}
    \caption{\textbf{Data Access Agent}. Inputs: user query $q$ and dataset root directory $R$. Output: the scientific data knowledge base $\mathcal{K}=\{D,T,C\}$, i.e., normalized data units and descriptors that serve as the shared input to downstream intent parsing, processing, and integration.}
    \label{fig:data_exploration}
\end{figure}

As shown in Figure~\ref{fig:data_exploration}, the Data Access Agent takes as input the user query $q$ and the dataset root directory $R$, and outputs a scientific data knowledge base
\begin{equation}
(q, R)\;\longrightarrow\;\mathcal{K} = \left\lbrace D,\, T,\, C \right\rbrace,
\end{equation}
where $D$ denotes a data lake, $T$ a tool lake, and $C$ a case lake. 
The produced knowledge base $\mathcal{K}$ will be consumed by following agents (Intent Parsing Agent, Data Processing Agent, and Data Integration Agent) as a unified dataset representation for planning and execution.

The data lake stores normalized dataset entities and their descriptions:
\begin{equation}
D = \left\lbrace \left[ D_i,\, l_{D_i} \right] \right\rbrace,
\qquad
D_i = \left\lbrace \left[ d_j,\, l_{d_j} \right] \right\rbrace,
\end{equation}
where each dataset $D_i$ is represented as a collection of data units $d_j$ paired with unit-level descriptors $l_{d_j}$, while $l_{D_i}$ provides a dataset-level description. 
A data unit is deliberately constrained to a single modality, a single spatiotemporal scope, and a single target object (or entity of interest). This constraint makes data units atomic and composable, and it provides a stable abstraction for downstream operations such as indexing, retrieval, validation, and pipeline synthesis. The descriptors $l_{D_i}$ and $l_{d_j}$ constitute the normalized metadata layer of the ontology, capturing modality tags, structural schema, spatiotemporal indices when available, provenance, quality indicators, and summary statistics in a format that is consistent across domains.

The tool lake models the processing ecosystem in the same normalized manner:
\begin{equation}
T = \left\lbrace \left[ T_i,\, l_{T_i} \right] \right\rbrace,
\qquad
T_i = \left\lbrace \left[ t_j,\, l_{t_j} \right] \right\rbrace,
\end{equation}
where tool $t_j$ is defined as a functionally complete module, and $l_{t_j}$ specifies its capability, input--output contract, dependencies, runtime constraints, and applicable modalities. 
By storing tools as first-class entities with explicit descriptors, SciDataCopilot can retrieve and compose tools based on task requirements and data-unit properties, rather than relying on hard-coded scripts or informal conventions. 
The collection-level descriptor $l_{T_i}$ further summarizes a tool set, supporting higher-level routing decisions and compatibility checks during planning.

The case lake captures reusable solutions that connect scientific intent, data, and tools:
\begin{equation}
C = \left\lbrace C_i \right\rbrace,
\qquad
C_i = \left\lbrace
l_{c_i},\,
\left[d_j,\, l_{d_j},\, (t_{j_1}, t_{j_2}, \ldots, t_{j_k})\right],\,
l_G,\,
(t_{G_1}, t_{G_2}, \ldots, t_{G_m})
\right\rbrace,
\end{equation}
where case $C_i$ contains a case description $l_{c_i}$ that records the scientific problem statement and the corresponding processing strategy. 
It also includes a set of data units, where each unit is associated with a tool sequence $(t_{j_1}, \ldots, t_{j_k})$ that forms a unit-level processing pipeline. 
Finally, the case defines an integration description $l_G$ and an integration tool sequence $(t_{G_1}, \ldots, t_{G_m})$ that specifies how multiple processed units are consolidated into coherent outputs. 
Under this formulation, the case lake serves as an explicit interface between data representations and executable tools: each case encodes procedural knowledge as a compositional graph over typed data units and typed tools, supporting case retrieval, pipeline construction, and traceable reuse across tasks.

The database design $\mathcal{K}=\{D,T,C\}$ provides a unified and comprehensive information platform that supports data association, integration, and sharing. 
By grounding all entities in normalized descriptors and explicitly modeling the links among datasets, atomic data units, tools, and reusable cases, SciDataCopilot establishes an ontology-oriented substrate on which downstream agentic planning and execution can operate reliably across heterogeneous scientific domains.

\subsubsection{Methodology of Data Access Agent}

\noindent\textbf{Recursive Exploration.}
Given a specified dataset root directory, the Data Access Agent performs a recursive scan to produce a complete inventory of files, and an index of relative paths capturing the dataset organization. 
This step provides a first-pass structural view of the dataset, including directory patterns, file groupings, and extension-based modality partitions.

\noindent\textbf{Type Recognition, Script Reuse, and an Execute--Reflection Loop.}
For each discovered asset, the agent first performs file type recognition, for example, \texttt{pkl}, \texttt{npy}, and \texttt{pt} for tensor data format, \texttt{csv} and \texttt{xlsx} for categorical table format, and \texttt{json}, \texttt{fasta} for sequence files. 
The system then prioritizes reusing previously validated exploration scripts for the same file type or similar structural patterns. 
If reuse fails or no reusable logic exists, the agent invokes a large language model to synthesize a new exploration script.
Crucially, exploration code is not assumed to work on the first attempt. 
Instead, the agent executes the script, captures runtime errors, and enters an automatic execute--error--reflection loop with a bounded iteration budget until it converges to a runnable parser and summarizer. 
This closed-loop execution makes dataset perception operationally robust and reduces reliance on manual debugging.

\noindent\textbf{Reproducible Data Unit Summaries and Traceable Artifacts.}
Once data access succeeds, the agent persists the results to disk, including the data unit identifier $d_j$ (e.g., file path, asset key, modality tag), and the corresponding description $l_{d_j}$.
The description $l_{d_j}$ includes essential statistical and structural summaries such as tensor shapes, field lists, column lists, data types, value ranges, missingness patterns, and representative samples. 
In addition, the agent stores the script used to generate the summary, enabling reproduction and verification.

\noindent\textbf{From Exploration Outputs to a Prototype Scientific Data Ontology.}
After exploring the data directory, the system turns what it discovers into structured information that later workflow stages can directly use. 
It collects metadata from many different files and formats, then cleans and standardizes it into a single dataset-level description, denoted as $l_{D_i}$. 
This description captures the essentials of the dataset---what kinds of data it contains, how fields are organized, what the basic statistics look like, whether there are plausible time or space indices, how the data might be split into training and evaluation sets, and what semantic cues can be inferred from folder structures and filenames. 
The resulting summary is stored in $l_{D_i}$ and serves as machine-readable context for the next step, such as intent parsing and pipeline planning.

\noindent
\textbf{Interface to Downstream Agents.}
In summary, the Data Access Agent grounds the entire system by producing $\mathcal{K}$ as a reusable, inspectable knowledge base. 
Downstream, the intent parsing agent queries $\mathcal{K}$ to select relevant data units and produce an intent-level plan; the processing agent executes these plans over the selected units; and the integration agent consolidates intermediate outputs into the final dataset-level deliverable.

\subsection{Intent Parsing Agent}
\label{subsec:parse}

\begin{figure}[t]
    \centering
    \includegraphics[width=1.0\textwidth]{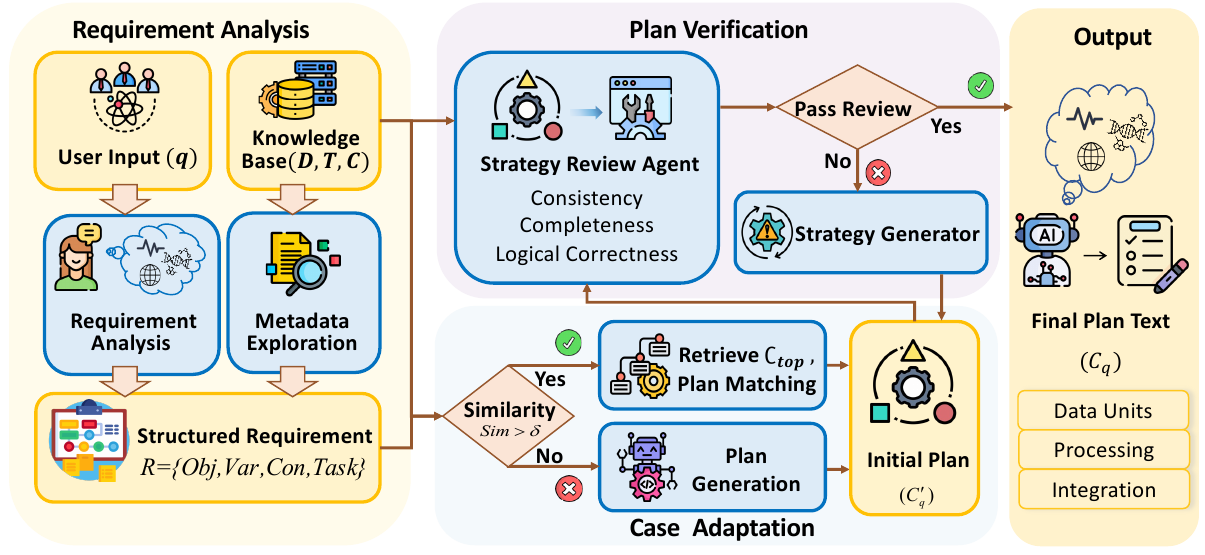} 
    \caption{\textbf{Intent Parsing Agent}. Leveraging the knowledge base $\mathcal{K}=\{D,T,C\}$, the agent processes user requirements $q$ and generates an executable plan $C_q$ through requirement analysis, case adaptation or generation, and iterative strategy review.}
    \label{fig:inter_parse}
\end{figure}

Building on the ontology-oriented knowledge base $\mathcal{K}=\{D,T,C\}$ in Section~\ref{subsec:access}, the intent parsing module in this section uses the normalized descriptors of data units, tools, and cases as semantic anchors to translate a natural-language query $q$ into a structured planning set $C_q$ with verifiable data units selections and executable processing strategies. The overall workflow is shown in Figure~\ref{fig:inter_parse}. Specifically, given a user-provided requirement description or scientific question $q$ and a knowledge base $\mathcal{K}=\{D,T,C\}$, the goal is to transform ambiguous intents into structured data units and corresponding processing strategies. The system automatically generates a planning set $C_q$ containing $n$ data units, which can be formally expressed as:
\begin{equation}
q
\xrightarrow{ \{D, T, C\} }
C_q .
\end{equation}
Here, $d_{q_k}$ denotes the $k$-th data unit relevant to the user query $q$; $l_{q_k}$ denotes the processing strategy for that data unit; and $l_{q,G}$ denotes the joint strategy for integrating multiple data units. The final output is a pre-executable textual plan, which can be subsequently grounded into concrete scripts and workflows by an executor. To facilitate the systematic translation of ambiguous requirements into concrete planning, our methodology adopts a cognitive framework comprising three sequential stages: requirement analysis, case retrieval and adaptation, and strategy generation and verification.

\subsubsection{Scientific Data Requirement Analysis}

The primary challenge in data analysis lies in establishing a semantic mapping between natural language and the knowledge base. This module parses user inputs via intelligent agents to extract core research variables and constraints, and forms a structured requirement representation $R=\{Obj, Var, Con, Task\}$,corresponding to the research object, key variables, constraint conditions (e.g., temporal granularity and quality thresholds), and the type of analytical task, respectively. During this process, the system conducts a lightweight metadata scan over $D$ to ensure that the extracted variables are physically available or can be constructed. For example, given the user input `\emph{Exploring the growth mechanism of plants under microgravity}' , the system derives:
\begin{equation}
    R = \begin{cases}
        Obj = \text{``plant''} \\
        Var = \text{``growth performance, growth mechanism''} \\
        Con = \text{``microgravity''} \\
        Task = \text{``exploratory study, mechanism analysis''}
    \end{cases}
\end{equation}
The structured requirement representation $R$ provides semantic anchors for subsequent retrieval in the case lake.

\subsubsection{Case Retrieval and Adaptation}
\label{sec:case_retrieval}

To reduce hallucinations induced by automatic planning and to improve planning quality and reproducibility, we introduce a case lake $C$ as prior solutions. The case lake consists of historical scientific questions and their corresponding data-processing solutions, formalized as:
\begin{equation}
    C=\{C_i\}, \quad C_i=\{l_{c_i},[d_j,l_{d_j},(t_{j_1},t_{j_2},\ldots,t_{j_k})],l_G,(t_{G_1},t_{G_2},\ldots,t_{G_k})\}.
\end{equation}
In this formulation, $C_i$ denotes the $i$-th case in the repository $C$, and $l_{c_i}$ denotes the semantic description of case $C_i$, covering both the natural-language scientific problem and the corresponding solution. The tuple $[d_j, l_{d_j}, (t_{j_1}, t_{j_2}, \ldots, t_{j_k})]$ records the data unit $d_j$ used in the case, the processing strategy $l_{d_j}$ applied to that data unit, and the associated tool sequence $(t_{j_1}, t_{j_2}, \ldots, t_{j_k})$. In addition, $l_G$ together with $(t_{G_1}, t_{G_2}, \ldots, t_{G_k})$ specifies the cross-unit integration strategy and the corresponding integration pipeline.

Based on this structure, we propose a hybrid inference mechanism comprising retrieval, adaptation, and generation phases:
\begin{itemize}
    \item \textbf{Case retrieval:} We compute the similarity $Sim(R,l_{c_i})$ between the user requirement $R$ and the semantic description $l_{c_i}$ in the case library. If a high-confidence case $C_{top}$ satisfies $Sim > \delta$, it is retrieved as a `\emph{ seed solution }' .
     \item \textbf{Plan adaptation:} If $C_{top}$ is successfully retrieved, the system performs differential analysis. Since new user questions often involve slight changes in variables or conditions (e.g., replacing `\emph{mouse}'  with `\emph{plant}', or replacing `\emph{regional environment}'  with `\emph{growth mechanism}' ), the system uses agents to map the original research object in $C_{top}$ to the new object in $R$, thereby revising the processing strategy $l_{d_j}$ and the integration strategy $l_G$ to produce an adapted plan.
      \item \textbf{Plan generation:} If no relevant case exists in the repository ($Sim < \delta$), the system falls back to a multi-step reasoning–driven generation mode. It infers the required data units from the intent logic in $R$, and calls large models via agents to synthesize domain knowledge to derive corresponding processing and integration strategies.
\end{itemize}

\subsubsection{Generation and Verification of Scientific Data-Processing Plans}

Subsequent to intent parsing and case adaptation, the system advances to the final phase of plan synthesis. To guarantee that the resulting plans are both logically sound and physically executable, we establish a closed-loop mechanism comprising generation, review, and revision steps. This iterative process is orchestrated jointly by a Plan Generator and a Plan Reviewer Agent.

\textbf{Initial Plan Generation.} In this phase, leveraging the adapted case derived in Section~\ref{sec:case_retrieval}, the system instantiates a natural-language plan tailored to the user's specific query. This plan is formally defined as:
\begin{equation}
    C_q = \{(d_{q_1},l_{q_1}), (d_{q_2},l_{q_2}), \ldots, (d_{q_n},l_{q_n}), l_{q,G}\}.
\end{equation}
The plan can be decomposed into three levels of content: $d_{q_i}$ denotes the extracted data units, explicitly specifying the $n$ independent data units required to address the target problem; $l_{q_i}$ denotes the per-unit processing strategy, which defines the cleansing and preprocessing logic to be applied to each corresponding data unit; and $l_{q,G}$ denotes the integration strategy, which formalizes the relationships among data units and the associated integration logic.

\textbf{Plan Review and Optimization.} Given that the initial plan may be prone to logical lacunae or resource unavailability, we deploy a strategy-review agent to conduct a multi-dimensional assessment of the generated plan $C_q$, utilizing the structured requirement $R$, the data lake $D$, and the tool repository $T$ as references. This validation process encompasses three critical dimensions: first, \textit{requirement alignment}, which verifies whether the generated plan is consistent with the user’s input requirement $q$ and detects off-topic responses, concept substitution, or scope drift (e.g., if the user requests an analysis of the ` \emph{plant growth mechanism} ' but the plan instead focuses on the ` \emph{root-system growth mechanism} ' or only discusses a single submodule, it is deemed misaligned); second, \textit{coverage completeness}, which verifies whether the strategy exhaustively covers the key variables and constraints defined in $R$ (e.g., if the user requests data on ` \textit{plant growth mechanism} ' but the plan only provides data pertaining to ` \textit{root growth mechanism} ', the plan is deemed incomplete); and third, \textit{logical correctness}, which evaluates the rationality of the integration strategy $l_{q,G}$ to preclude structural errors, such as the invalid merging of data possessing mismatched temporal granularities.

\textbf{Iterative Revision and Plan Finalization.} If the strategy-review agent detects potential issues, it generates natural-language feedback specifying concrete error causes. The plan generator then leverages the contextual learning ability of large models to locally revise $C_q$. This process may iterate for multiple rounds until the strategy-review agent approves the plan or a maximum iteration limit is reached. The final output is a rigorously validated, executable planning text.

\subsection{Data Processing Agent}
\label{subsec:process}
\begin{figure}[t]
    \centering
    \includegraphics[width=1.0\textwidth]{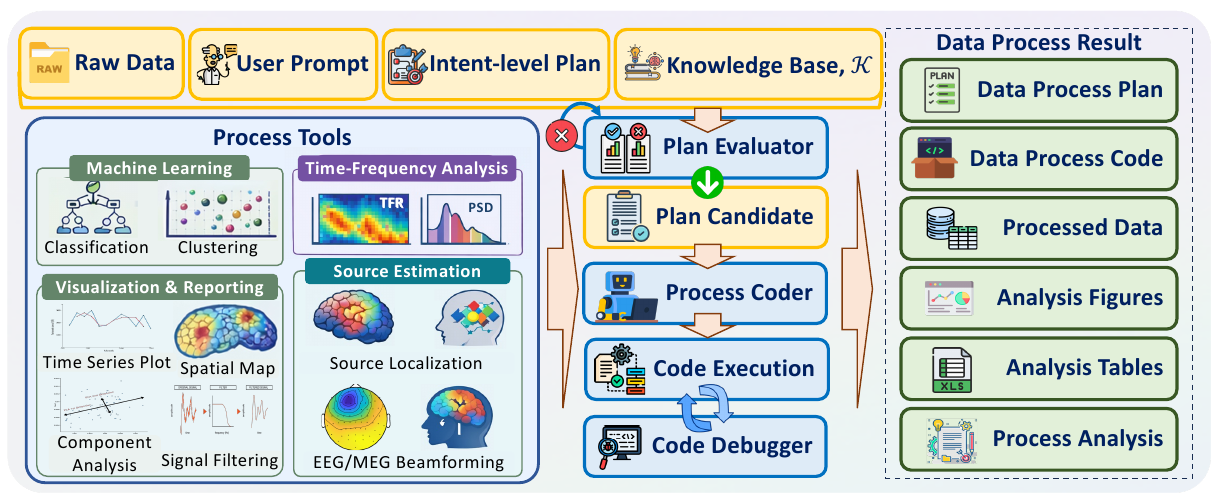}
    \caption{\textbf{Data Processing Agent}. Given the user query $q$, the selected raw data units, and an intent-level processing plan, the agent executes reproducible pipelines to produce processed data, visualizations, and analysis summaries for downstream integration.}
    \label{fig:data_process}
\end{figure}

As illustrated in Figure~\ref{fig:data_process}, the Data Processing Agent consists of five core modules::
(1) \textbf{Plan refinement} revises the initial plans from the intent parsing stage into feasible, executable pipeline specifications under data and resource constraints, focusing on \emph{what to do} rather than producing long code.
(2) \textbf{Plan checking} performs a technical review of the refined pipelines for correctness and feasibility, either approving them or returning concrete revision suggestions and routing the workflow back to refinement when necessary.
(3) \textbf{Code synthesis} produces a complete runnable program based on the approved refined plans and the current error context, and it versions each revision to enable post-mortem analysis and reproducibility.
(4) \textbf{Execution} runs the generated program, captures runtime outputs and error traces, writes logs into the experiment workspace, and feeds failures back into the shared state to trigger automatic repair.
(5) \textbf{Analysis} produces the scientific interpretation report when execution succeeds; otherwise, after repeated failures, it generates an auditable termination report that summarizes the failure reason and the iteration trace.

\subsubsection{Processing Pipeline}

In Figure~\ref{fig:data_process}, given the data unit plan $\{(d_{q_i}, l_{q_i})\}_{i=1}^n$ derived from the user query $q$ (and optionally a global integration strategy $l_{q,G}$), the Data Processing Agent executes each strategy $l_{q_i}$ over the corresponding data unit $d_{q_i}$ and produces intermediate, unit-level outputs that are ready for downstream consolidation.

Scientific data processing is highly dependent on domain knowledge and tooling ecosystems. 
Different disciplines exhibit substantial variation in pre-processing conventions, statistical assumptions, feature construction, and evaluation metrics. 
Moreover, real-world data often suffer from missing values, outliers, distribution shifts, and sampling biases, making fixed template scripts brittle over time. 
Therefore, the system emphasizes two principles, i.e., controllable domain reasoning and executable closed loop.
The principle of controllable domain reasoning is to leverage large-model reasoning for pipeline orchestration while keeping the process auditable and bounded by validation.
In contrast, executable closed loop is to treat code generation as an executable, verifiable, and self-repairing loop rather than a one-shot text output.

Formally, the agent takes as input the user request $q$, the intent-parsed processing plan, and the shared knowledge base $\mathcal{K}=\{D,T,C\}$ constructed in the data access stage:
\begin{equation}
\left[q,\; \mathcal{P}=\{(d_{q_i},l_{q_i})\}_{i=1}^n,\; \mathcal{K}\right]\;\longrightarrow\;\left[\mathcal{O},\; \mathcal{V},\; \mathcal{A}\right].
\end{equation}
Here, $\mathcal{O}=\{d_{o_i}\}_{i=1}^n$ denotes the processed data units, $\mathcal{V}$ denotes the generated visualization set (plots/tables for diagnosis and reporting), and $\mathcal{A}$ denotes concise analysis summaries (e.g., key statistics, sanity-check outcomes, and metric reports). 
These outputs form the intermediate deliverables that will be assembled by the downstream DataIntegrationAgent according to $l_{q,G}$.

Following the data processing specification produced in the upstream planning stage, our system aims to automatically select appropriate data processing tools, automatically refine a reproducible processing pipeline, automatically generate executable data processing code, and automatically optimize execution efficiency through iterative debugging and validation. 
In addition, the system accounts for practical characteristics of scientific data processing, including unit-level heterogeneity, dependency-aware scheduling, and parallel execution over independent data units to improve throughput under limited compute and I/O constraints.

Let the intent parsing agent extract a set of required data units
\begin{equation}
\mathcal{D}_s = \left\lbrace d_{s_1}, d_{s_2}, \ldots, d_{s_n}\right\rbrace,
\end{equation}
and produce a processing strategy for each unit
\begin{equation}
\mathcal{L}_s = \left\lbrace l_{s_1}, l_{s_2}, \ldots, l_{s_n}\right\rbrace.
\end{equation}
The intent parsing agent further provides an initial processing plan for each unit, serving as a starting point for downstream refinement,
\begin{equation}
\mathcal{P}_s^{(0)} = \left\lbrace p_{s_1}^{(0)}, p_{s_2}^{(0)}, \ldots, p_{s_n}^{(0)} \right\rbrace.
\end{equation}
Given a domain knowledge base that includes a case lake of prior pipelines, tool usages, and best practices, the data processing agent refines each initial plan $p_{s_i}^{(0)}$ into an executable unit-level pipeline specification $p_{s_i}$, and composes, for each strategy $l_{s_i}$, a tool-sequence as a pipeline of the form
\begin{equation}
\mathcal{T}_{s_i} = \left( t_{s_i,j_1}, t_{s_i,j_2}, \ldots, t_{s_i,j_k} \right),
\end{equation}
and executes these pipelines to transform the extracted data units into processed outputs
\begin{equation}
\mathcal{D}_o = \left\lbrace d_{o_1}, d_{o_2}, \ldots, d_{o_n}\right\rbrace.
\end{equation}
In addition to $\mathcal{D}_o$, the agent produces visualization artifacts (e.g., diagnostic plots and summary tables) and lightweight analysis summaries that support quality control and technical reporting.

We implement the scientific data processing agent as an explicit state-machine workflow, where each step has a clear responsibility and the system can decide, based on intermediate results, which step to run next. 
The end-to-end workflow is organized into five phases. 
First, the agent retrieves relevant prior cases from its knowledge base and takes the initial plans from the intent parsing stage as inputs. 
Second, it performs plan refinement, where it revises these initial plans into executable pipeline specifications by resolving tools choices, enforcing input--output compatibility, and incorporating feasibility constraints such as data volume, memory budget, and runtime limits. 
Third, once the refined plans are accepted, the agent generates a complete, runnable program that implements the proposed pipelines. 
Fourth, the system enters a debugging stage in which it reacts to warnings and runtime errors by repairing the program. 
Finally, it executes the program, collects execution feedback, and iterates until the run succeeds or the process terminates. 
Throughout this process, the agent maintains a single shared runtime state that records the evolving plans, generated code, execution results, and intermediate artifacts.

For each candidate processing plan, the agent searches a repository of past solutions to find similar instances and uses them to assemble an appropriate sequence of tools and operations. 
The resulting pipeline is checked for feasibility before any code is produced. 
If the plan is not viable, the agent requests model-based critique and uses the feedback to refine the plan. 
Once the plan is deemed valid, the code-generation component produces an end-to-end implementation and saves it as a versioned artifact within a dedicated experiment workspace, enabling later inspection and reuse. 
To reflect practical data-processing characteristics, the agent identifies independent units and schedules them for parallel execution when their dependencies allow, while keeping shared resources (e.g., disk bandwidth and GPU memory) bounded through concurrency limits and staged execution.

\subsubsection{Self-Repairing Execution Loop}

We do not assume that generated programs will run correctly on the first attempt. 
Instead, program execution and program synthesis form a closed feedback loop. 
The executor runs the current program and captures the full runtime outputs, including standard output and error traces. 
These signals are written back into the shared state and then used to guide a repair step, where the code-generation component rewrites a complete runnable program conditioned on both the intended refined plans and the latest runtime failure. 
In effect, the system moves beyond producing code as static text and treats code generation as an iterative engineering process driven by real execution feedback, which improves robustness in practical environments.

When execution succeeds, the agent produces an analysis suitable for technical reporting, including scientific interpretation of the results, explanations of the computed metrics, and a concise conclusion summary. 
When repeated attempts fail to converge, the system does not fail silently. 
Instead, it produces a traceable failure report that records why the run terminated and preserves the iteration history, making the outcome auditable and actionable rather than opaque.

\subsubsection{Reproducibility and Traceability}

The system is designed to support reproducible engineering delivery.
Each run creates a timestamped experiment workspace that snapshots the refined pipeline plans, every program revision, execution logs, and the final report.
By connecting dataset-ingestion outputs (normalized dataset metadata in $\mathcal{K}$) with processing outputs (refined plans, implementations, visualizations, and reports), the agent forms a closed loop of data-grounded automation: dataset understanding informs planning, planning drives tool selection and implementation, and execution validates correctness.
This design supports scalable automation across heterogeneous scientific datasets while preserving traceability for debugging and review.

The runtime is governed by a unified agent state that stores all information required to coordinate the workflow.
It includes the user's request and task type, the dataset identifiers and normalized dataset description, the initial plans from the intent parsing stage and the current refined plans, the program implementation, iteration counters and recorded errors from both refinement and execution, and the final analysis outputs together with a success flag.
The state also tracks the current experiment workspace so that all artifacts created during a run are associated with a concrete, inspectable directory.


\subsection{Data Integration Agent}
\label{subsec:integrate}
The preceding data processing stage produces a collection of processed data units
$\left\{ d_{o_1}, d_{o_2}, \ldots, d_{o_n} \right\}$.
The objective of the proposed \textbf{Data Integration Agent} is to consolidate these distributed and heterogeneous data units into a single data object $D_{\text{uni}}$ with coherent global semantics:
\begin{equation}
\{ d_{o_1}, d_{o_2}, \dots, d_{o_n} \}
\xrightarrow{ \{(t_{o,G_1}, t_{o,G_2}, \dots, t_{o,G_k})\} }
D_{\text{uni}} .
\end{equation}

\begin{figure}[t]
    \centering
    \includegraphics[width=1.0\linewidth]{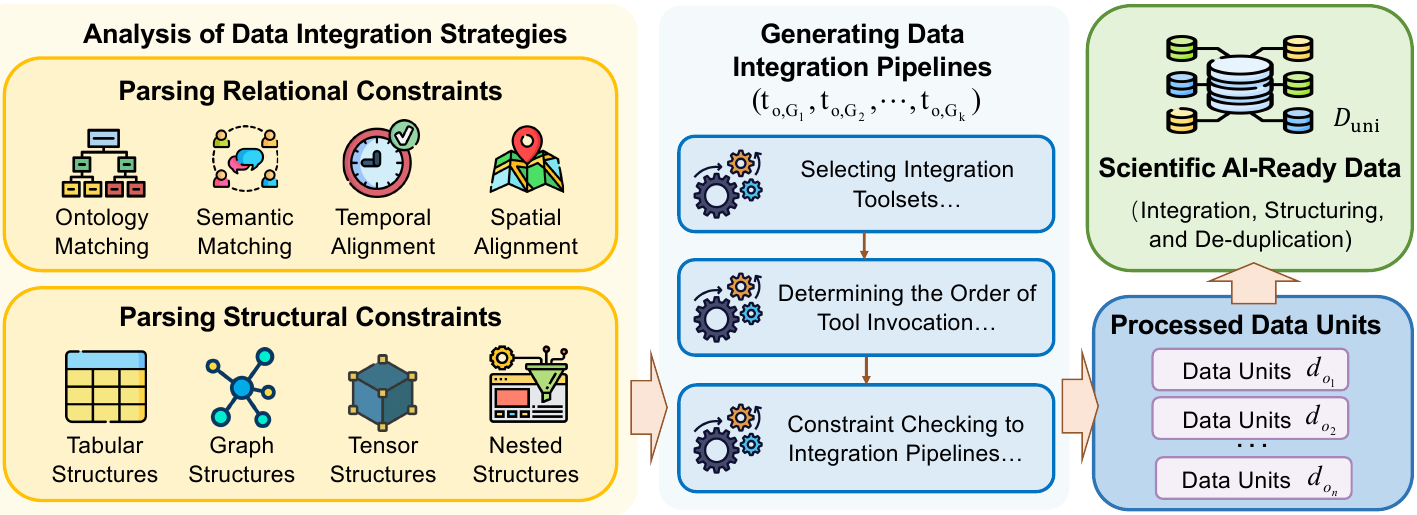}
    \caption{\textbf{Data Integration Agent}. The system operationalizes integration strategies by enforcing structural constraints, selecting validated tool sequences, and assembling distributed data units into a cohesive Scientific AI-Ready dataset.}
    \label{fig:data_integration}
\end{figure}

Here, the integration pipeline
$\left\{ (t_{o,G_1}, t_{o,G_2}, \ldots, t_{o,G_k}) \right\}$
is generated from an integration strategy $l_{o,G}$ produced during the upstream intent parsing and planning stage.
Importantly, $l_{o,G}$ is not an operational instruction sequence, but a joint \emph{relation--structure} constraint specification (Figure ~\ref{fig:data_integration}).
It encodes how data units should be related, aligned, and aggregated, as well as the desired global organization of the final dataset.
Each integration operator $t_{o,G_i}$ is selected from the tool knowledge base (tool lake, $T=\left\{ [T_i, l_{T_i}] \right\}$), where $T_i$ denotes an available integration tool and $l_{T_i}$ specifies its functional semantics and constraints.

\subsubsection{Analysis of Data Integration Strategies}

Given an integration strategy $l_{o,G}$, the goal of this stage is to transform high-level integration intents—typically expressed in natural language or abstract semantic descriptions—into a set of formal, explicit, and verifiable integration constraints:

\begin{equation}
l_{o,G} \;\longrightarrow\; 
\Gamma_G = \{ \gamma_1, \gamma_2, \ldots, \gamma_m \}.
\end{equation}

Each integration constraint $\gamma_k$ is defined as:
\begin{equation}
\gamma_k = (R_k, S_k),
\end{equation}
where $R_k$ specifies the type of relationships that must hold among the data units
$\left\{ d_{o_1}, d_{o_2}, \ldots, d_{o_n} \right\}$, and $S_k$ specifies the required structural form of the integrated output.

The relationship component $R_k$ may include, but is not limited to, ontology alignment, semantic correspondence, temporal synchronization, and variable or schema mapping. The structure component $S_k$ defines the expected representation of the integrated dataset, such as tabular structures, graph-based structures, or higher-order tensor representations.
Both $R_k$ and $S_k$ are inferred by the LLM based on the original user query $q$ and the global scientific intent. Specifically, the LLM determines which data units must be related, the nature of their relationships (e.g., semantic, temporal, or variable-level), and the structural requirements that the final dataset must satisfy in order to be Scientific AI-Ready.

\subsubsection{Generating Data Integration Pipelines}

Given the set of integration constraints
$\Gamma_G = \{ \gamma_1, \gamma_2, \ldots, \gamma_m \}$
and the tool lake $
T = \left\{ [T_i, l_{T_i}] \right\}$,
the system generates an integration tools sequence $\left\{ (t_{o,G_1}, t_{o,G_2}, \ldots, t_{o,G_k}) \right\}$.

Each selected integration tool $t_{o,G_i}$ must jointly satisfy all constraints defined in $\Gamma_G$, including both the required relationship types $R_k$ and the target structural forms $S_k$. The generation of the integration pipeline requires the LLM to reason over the following aspects:

\begin{enumerate}
    \item \textbf{Tool--constraint matching}: identifying which tools are capable of realizing specific relationship constraints $R_k$, and which tools can produce outputs conforming to the structural requirements $S_k$;
    \item \textbf{Sequential reasoning over tools}: determining a valid execution order of tools, for example, performing temporal alignment prior to variable mapping or aggregation;
    \item \textbf{Failure-aware backtracking}: if a selected tool fails to satisfy a constraint $\gamma_k = (R_k, S_k)$, the system must revise its decisions by reselecting alternative tools or restructuring the integration sequence.
\end{enumerate}

This process ensures that the resulting integration pipeline is both semantically sound and operationally feasible under the specified constraints. Finally, the generated ordered integration tool sequence $\left\{ (t_{o,G_1}, t_{o,G_2}, \ldots, t_{o,G_k}) \right\}$
is executed sequentially in the specified order over the processed data units$\left\{ d_{o_1}, d_{o_2}, \ldots, d_{o_n} \right\}$
to produce the final integrated dataset:
\begin{equation}
D_{\text{uni}} = \left\{ [d_i, l_{d_i}] \right\}.
\end{equation}

Here, each $d_i$ denotes a component of the integrated dataset, and $l_{d_i}$ captures its associated semantic annotations and structural metadata. The resulting $D_{\text{uni}}$ constitutes a coherent, globally aligned, and AI-Ready dataset that directly satisfies the original scientific intent expressed in the query $q$.
\section{Use Cases}
\label{sec:cases}
To assess the practical effectiveness of SciDataCopilot in real scientific workflows, we evaluated the framework on three use cases spanning multiple disciplines and data modalities.
Each case was executed end-to-end from natural-language instructions under consistent experimental settings, with outcomes compared against expert human workflows.
All experiments were conducted on a server equipped with Intel Xeon Platinum 8558 CPU, and the LLM used in the framework is GPT-5.2~\cite{GPT_5_2_2025}.
Evaluation focused on efficiency, output standardization, and robustness under realistic data and tooling constraints.

\subsection{Case 1: Automated Collection of Enzyme-Catalysis Data}
\label{subsec:protein_case}
\subsubsection{Problem Statement}

In many life science domains, emerging research problems lack standardized, task-ready datasets that can be directly used for downstream modeling and analysis. Enzyme catalysis studies, in particular, often require the joint availability of enzyme sequences, catalytic reactions, and chemically explicit substrate--product representations, which are rarely provided in a unified and machine-readable form.

As a consequence, data preparation typically relies on multi-stage and manually intensive workflows, including literature inspection, querying multiple heterogeneous databases, and extensive data cleaning and normalization. These fragmented processes limit scalability and reproducibility, and pose a substantial barrier to the rapid application of data-driven methods in enzyme catalysis research.

This section examines enzyme catalysis data acquisition as a representative use case, evaluating whether high-level natural language specifications can be reliably translated into structured, machine-readable enzyme catalysis datasets at scale.

\begin{figure}[t]
    \centering
    \includegraphics[width=1.0\linewidth]{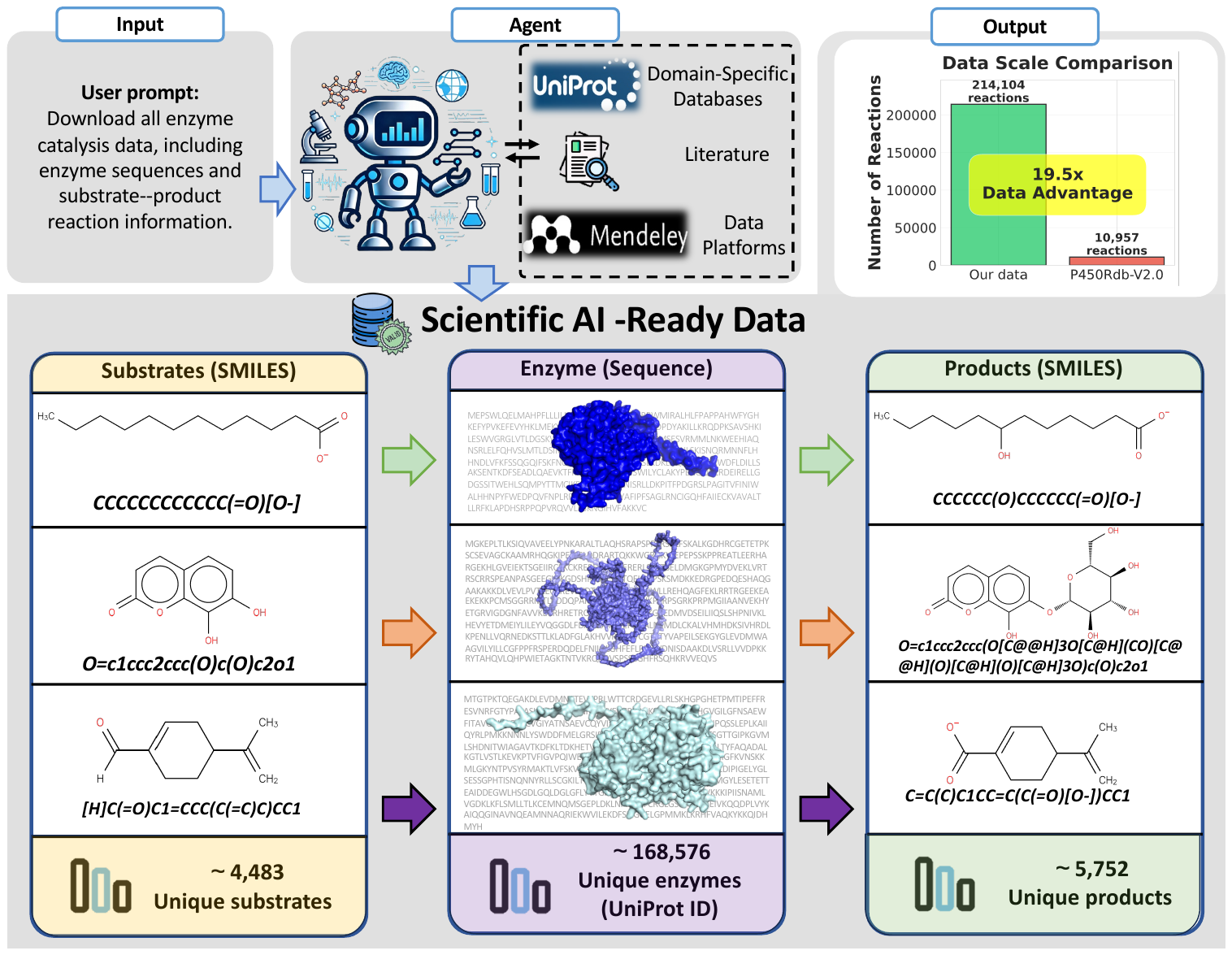}
    \caption{Case Study Diagram of an Automatically Curated Enzyme Catalysis Database Prepared by SciDataCopilot.}
    \label{fig:Case_protein_enzyme}
\end{figure}

\subsubsection{Experiment Setting}

In this case study, SciDataCopilot is applied to a large-scale protein knowledge base to evaluate its ability to automatically acquire and preprocess enzyme catalysis data from a natural language specification. The focus of the experiment is on assessing whether high-level scientific intents can be translated into structured, machine-learning-ready datasets without requiring users to manually interact with database schemas or query interfaces.

\begin{itemize}
    \item \textbf{User prompts}: Download all enzyme catalysis data, including enzyme sequences and substrate--product reaction information.
\end{itemize}

We consider enzyme catalysis as a representative task due to its reliance on heterogeneous information, including protein sequences, functional annotations, and chemically explicit reaction descriptions. The experimental input is a single natural language query requesting comprehensive enzyme catalysis data, including enzyme sequences and corresponding substrate--product reactions. This setting allows us to evaluate the end-to-end data acquisition capability of SciDataCopilot under minimal user intervention.

Additional query examples, such as protein family filtering or organism-specific data collection, are supported by the same framework but are not the primary focus of this evaluation. These prompts are exclusively formulated based on the user's requirements, as exemplified by the following instances.

\begin{itemize}
    \item User prompt 1: Collecting catalytically active proteins in humans.
    \item User prompt 2: Searching for RNA-binding proteins in mouse.
    \item User prompt 3: Extracting protein kinase sequences and substrate information.
    \item User prompt 4: Retrieving structural and functional data for DNA-binding proteins.
    \item User prompt 5: Collecting oxidoreductase enzymes and their catalytic reactions.
\end{itemize}

\subsubsection{Experimental Results}

We evaluate SciDataCopilot on the task of large-scale enzyme catalysis data acquisition and curation. Given a natural language specification, the system automatically collects and preprocesses enzyme-related information from protein knowledge bases, producing machine-learning-ready enzyme--reaction records with explicit substrate and product representations.

Using this pipeline, we construct a large-scale enzyme catalysis dataset comprising 214,104 structured reaction records (Figure~\ref{fig:Case_protein_enzyme}). Each record includes the enzyme sequence, a normalized catalytic reaction description, and corresponding substrate--product SMILES pairs. The resulting dataset supports a range of downstream applications, including enzyme specificity prediction, catalytic reaction prediction
, and science foundation models.

Figure~\ref{fig:Case_protein_enzyme} summarizes the overall composition of the curated dataset, illustrating enzyme family diversity, reaction coverage, and substrate--product statistics. The dataset spans thousands of distinct substrates and products and covers a broad spectrum of enzyme classes, indicating substantial chemical and functional diversity. Notably, the entire data collection and preprocessing process can be completed autonomously by SciDataCopilot within approximately five hours, requiring only minimal human inspection for final result validation.

We compare the curated dataset with P450Rdb-v2.0~\cite{zhang2024p450rdb}, a specialized database focused on cytochrome P450 enzymes. Our dataset contains 214,104 enzymatic reactions associated with 168,576 unique enzymes, 4,483 substrates, and 5,752 products, representing an approximately 19.5$\times$ increase in overall reaction scale relative to P450Rdb-v2.0, which includes 10,957 curated P450-specific reactions. 

This scale advantage reflects the broad enzyme coverage enabled by SciDataCopilot, encompassing diverse families such as methyltransferases, dehydrogenases, kinases, oxidoreductases, and transferases, and is well suited for data-driven modeling that benefits from large and heterogeneous training corpora. At present, the number of cytochrome P450 reactions is smaller than that in P450Rdb-v2.0 due to the current reliance on only UniProt-derived data sources; however, with the planned integration of literature-derived and public biochemical databases, we expect both the scale and quality of P450-related data to exceed existing domain-specific resources.

Overall, these results demonstrate that SciDataCopilot enables rapid, large-scale construction of enzyme catalysis datasets, providing a scalable foundation for continued expansion toward richer and more comprehensive biochemical coverage.

\subsubsection{Summary}

The enzyme catalysis case study illustrates that the proposed data curation paradigm is not inherently tied to a single database, schema, or biological domain. The same interaction pattern---specifying high-level scientific intent through natural language and obtaining structured, task-ready data---can be applied to other data acquisition scenarios that involve heterogeneous and loosely standardized scientific resources.

Beyond enzyme catalysis, this paradigm is applicable to a range of related tasks, such as collecting protein--ligand interaction data, aggregating structural and functional annotations across databases, or assembling organism-specific molecular datasets. More broadly, the case study suggests that automated data curation can serve as a unifying interface across protein databases, biochemical resources, and other structured scientific repositories.

These observations indicate that SciDataCopilot provides a generalizable mechanism for scalable data preparation, supporting reproducible dataset construction across diverse scientific domains with minimal manual effort.

\subsection{Case 2: End-to-end Neuroscience Data Analysis}
\label{subsec:eeg_case}
\begin{figure}[h]
    \centering
    \includegraphics[width=1.0\textwidth]{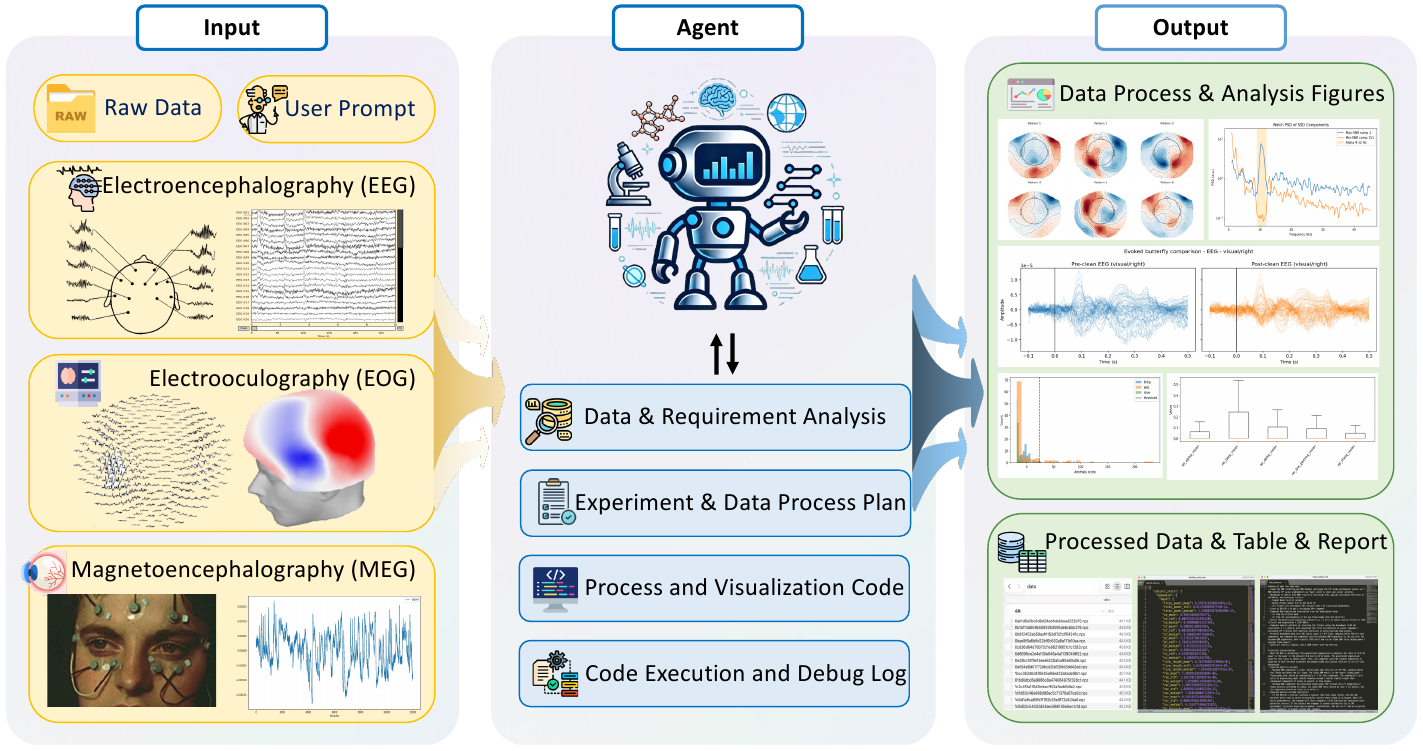}
    \caption{Case Study Diagram of the SciDataCopilot on Electroencephalography (EEG), Electrooculography (EOG), and Magnetoencephalography (MEG). }
    \label{fig:eeg_case}
\end{figure}

This use case focuses on automated analysis of electrophysiology recordings, including electroencephalography (EEG), electrooculography (EOG), and magnetoencephalography (MEG). In neuroscience labs, such data is valuable but notoriously difficult to process at scale. Teams often face fragmented toolchains, inconsistent parameter choices, and poor reproducibility. Our goal is to transform this workflow into a standardized, modular, and reproducible pipeline that can be executed consistently across datasets, research groups, and experimental stages.

\subsubsection{Problem Statement}
EEG and MEG analysis is a fundamental technique of modern neuroscience, but real-world practice still suffers from several long-standing obstacles. First, the data are highly heterogeneous: different acquisition systems produce different file formats, metadata conventions, channel definitions, and sampling rates, which makes it difficult to build a unified pipeline that works reliably across datasets. Second, workflows are often fragmented across multiple software packages and scripts, with hand-written glue code that is brittle, difficult to maintain, and prone to failure when datasets change. Third, key pre-processing and analysis choices, such as filter settings, artifact rejection thresholds, and independent component analysis configurations, are frequently tuned by experience, which introduces operator bias and reduces reproducibility. Fourth, validation is often incomplete or inconsistent: researchers may inspect plots, but a unified framework that systematically checks statistical quality, physiological plausibility, and interpretability is often lacking. Finally, deliverables are commonly scattered across different folders and tool outputs, making collaboration, auditing, and re-running experiments unnecessarily time-consuming. To address these issues, we aim to transform expert-driven ``step-by-step'' electrophysiology processing into a standardized, modular, and reproducible workflow that links raw recordings to validated and auditable scientific outputs.

\subsubsection{Experimental Setting}
SciDataCopilot converts an expert-driven workflow into an automated end-to-end pipeline that begins from the user's research intent and ends with a structured set of scientific outputs.
The framework interprets an analysis request and maps it to a consistent configuration, loads and normalizes raw recordings from different devices and file formats, applies standardized preprocessing, runs a task-specific analysis module, performs multi-view validation, and exports processed data, figures, metrics, and a report in a traceable directory structure.
We evaluate it in representative electrophysiology scenarios using commonly used public datasets and established analysis practices.
The agent operates on recordings originating from multiple acquisition ecosystems, covering widely used electrophysiology formats such as CTF datasets, MNE-formatted files, and large-scale clinical EEG segment collections.
Concretely, we test four sub-tasks that frequently appear in neuroscience workflows:
(i) alpha-band activity extraction from MEG with spectral and spatial plausibility checks (Task~1),
(ii) ocular artifact correction on EEG/MEG with pre/post validation (Task~2),
(iii) ICA decomposition with numerical validity and interpretability checks (Task~3),
and (iv) large-scale clinical EEG preprocessing and feature extraction with seizure-related anomaly screening (Task~4).
To address both practicality and scientific validity, we evaluate along two axes:
\textbf{(i) Efficiency:} end-to-end completion time from an analysis request to a fully packaged deliverable (processed data, figures, metrics, and a report);
\textbf{(ii) Effectiveness:} task-specific outcome metrics paired with qualitative diagnostic evidence.
For Task~1, we report an alpha-to-flank spectral SNR and inspect spatial patterns and PSD comparisons.
For Tasks~2--3, we report quantitative model-fit and stability indicators when available (e.g., variance explained, runtimes, convergence/anomaly statistics) and provide auditable diagnostics (pre/post plots, spatial weight maps, component topographies, and logs).
For Task~4, we report dataset-level integrity metrics (valid segment rate, sampling-rate and channel consistency), preprocessing actions (artifact rejections and ICA removal statistics), and anomaly-screening performance measured by confusion counts and class-wise anomaly rates, together with exported per-file preprocessing logs for auditability.
Across all tasks, the agent records configurations, intermediate products, logs, and a final report so that the entire run can be reproduced in a self-contained workspace.
To provide a realistic baseline, we additionally recruited a human tester with electrophysiology processing experience to execute the same four tasks using standard toolchains and to deliver comparable outputs.
We report side-by-side comparisons between SciDataCopilot and the human baseline in terms of task-specific quantitative metrics and end-to-end completion time (Table~\ref{tab:eeg_all_tasks_summary}).

\subsubsection{Experimental Results}
Across the representative tasks, the agent produces complete end-to-end deliverables and demonstrates reliable behavior under heterogeneous inputs.

\textbf{Task~1 (MEG alpha extraction).}
Using the public FieldTrip CMC dataset (CTF MEG), the agent loads \texttt{SubjectCMC.ds}, crops to $[50,110]$\,s, resamples to 250\,Hz, and operates on 151 CTF axial gradiometers.
It performs SSD with a signal band of 9--12\,Hz and flanking noise bands of 8--9\,Hz and 12--13\,Hz (OAS-regularized covariance), together with a practical preprocessing backbone (50/100\,Hz notch and 1--45\,Hz zero-phase FIR band-pass) to stabilize estimation.
The extracted component achieves an alpha-to-flank power ratio of $23.53$ (13.72\,dB), while the lowest-SNR component yields $0.53$ (-2.78\,dB), indicating a strong max/min separation of $44.6\times$.
Qualitatively, exported PSD comparisons and topographic patterns show a sharper alpha peak for the selected component; CTF triplet sensors are handled via position-aware deduplication, which is recorded for interpretability.
For comparison, the human tester achieves a similar SNR (max/min) of $23.71$ (14.11\,dB) / $0.52$ (-2.59\,dB), but requires a longer end-to-end time of 31.9\,min (vs.\ 9.2\,min for the agent).

\textbf{Task~2 (EOG regression).}
On the MNE sample dataset, the agent identifies stimulus events (320 total; \{\texttt{visual/left}:3, \texttt{visual/right}:4\}), applies a 0.3\,Hz high-pass filter, and removes ocular artifacts by regressing \texttt{EOG 061} with a short-lag design (within $\pm$100\,ms) and ridge regularization.
Quantitatively, the regression explains a substantial fraction of ocular-driven variance (mean variance explained $0.8638$, median $1.0000$).
Qualitatively, the deliverable workspace contains before/after evoked comparisons and regression-weight topomaps that support plausibility checks (stronger coupling to frontal/peri-orbital sensors) and preservation of evoked morphology with reduced blink contamination.
Optional steps requiring external resources (e.g., Maxwell filtering) are treated as best-effort modules: missing files are logged while the pipeline continues with validated alternatives.
In contrast, the human baseline attains mean/median variance explained of $0.8614/1.0000$, while the end-to-end time increases to 40.5\,min (vs.\ 11.3\,min for the agent).

\textbf{Task~3 (ICA decomposition).}
The agent fits ICA on MEG-only signals from a 0--60\,s segment with a 1--30\,Hz zero-phase FIR band-pass filter, standardizing to 20 components and recording runtime/stability indicators.
FastICA converges in 62 iterations (0.43\,s), while Infomax and extended-Infomax complete successfully (0.69\,s and 0.75\,s; both reaching 500 iterations).
Picard is automatically skipped due to a missing optional dependency, detected and logged rather than breaking execution.
Component maps and summaries are exported for interpretability checks and artifact-like pattern inspection.
Under the same setting, the human run reports longer ICA backend runtimes (FastICA 0.7\,s; Infomax 2.1\,s; Ext-Infomax 3.2\,s; Picard 0.9\,s, all with the same iteration caps), and the end-to-end completion time is 35.4\,min (vs.\ 13.8\,min for the agent).

\textbf{Task~4 (TUSL large-scale EEG preprocessing and anomaly screening).}
The agent processes 300 TUSL EEG segments under a unified configuration (band-pass 0.5--40\,Hz; target SR 250\,Hz), yielding 296 valid segments with 4 rejections by artifact criteria.
Sampling rate is fully consistent (unique 250\,Hz), while channels vary mildly (min 21, median 23, max 23); per-file preprocessing logs record dropped channels, removed ICA indices, and bad-sample fractions for auditability.
Features (PSD and differential entropy over standard frequency bands) are computed using 2.0\,s windows with 0.5 overlap, and a global anomaly threshold (score 23.19) is used for seizure-related screening.
At this operating point, class-wise anomaly rates are $5.05\%$ (class~0), $14.0\%$ (class~1), and $0\%$ (class~2), with confusion counts $\mathrm{TP}=14$, $\mathrm{FP}=5$, $\mathrm{FN}=86$, and $\mathrm{TN}=191$.
This indicates a conservative screening behavior (low FP but high FN), and the agent makes the trade-off explicit by exporting dataset-level summaries and per-file audit traces for threshold tuning and clinical review.
Compared with the agent, the human tester processes 298/300 files successfully (2 rejected) and obtains confusion counts $\mathrm{TP}=20$, $\mathrm{FP}=11$, $\mathrm{FN}=62$, $\mathrm{TN}=203$, but the end-to-end time is substantially longer at 51.9\,min (vs.\ 15.6\,min for the agent).


\begin{table}[t]
\centering
\footnotesize
\renewcommand{\arraystretch}{1.12}
\setlength{\tabcolsep}{4pt}
\caption{Compact quantitative summary across four electrophysiology tasks, comparing SciDataCopilot against human researchers in terms of core metrics, runtime, and auditable artifacts.}
\label{tab:eeg_all_tasks_summary}
\begin{tabularx}{\linewidth}{@{}>{\RaggedRight\arraybackslash}p{2.7cm}
>{\RaggedRight\arraybackslash}X
>{\RaggedRight\arraybackslash}X
>{\RaggedRight\arraybackslash}X
>{\RaggedRight\arraybackslash}p{2.6cm}@{}}
\toprule
\textbf{Task} & \textbf{Dataset / Setting} & \textbf{SciDataCopilot} & \textbf{Human researchers} & \textbf{Audit artifacts} \\
\midrule

Task~1: Alpha extraction (SSD) &
FieldTrip CMC (CTF MEG); 250\,Hz; 151 chans &
\textbf{Metric:} SNR (max/min) $23.53$ (13.72\,dB) / $0.53$ (-2.78\,dB) \newline
\textbf{Time:} 9.2\,min &
\textbf{Metric:} SNR (max/min) $23.71$ (14.11\,dB) / $0.52$ (-2.59\,dB) \newline
\textbf{Time:} 31.9\,min &
PSD comparisons; topomaps; report \\

\addlinespace[2pt]

Task~2: EOG regression &
MNE sample; 0.3\,Hz high-pass; regressor \texttt{EOG 061} &
\textbf{Metric:} Var. explained mean $0.8638$, median $1.0000$ \newline
\textbf{Time:} 11.3\,min &
\textbf{Metric:} Var. explained mean $0.8614$, median $1.0000$ \newline
\textbf{Time:} 40.5\,min &
Pre/post evoked; weight topomaps; model \\

\addlinespace[2pt]

Task~3: ICA decomposition &
MNE sample; MEG-only; 0--60\,s; 1--30\,Hz band-pass; 20 comps &
\textbf{Runtime (s / iters):} FastICA 0.43/62; Picard 0.83/500 \newline
Infomax 0.69/500; Ext-Infomax 0.75/500 \newline
\textbf{Time:} 13.8\,min &
\textbf{Runtime (s / iters):} FastICA 0.7/62; Picard 0.9/500 \newline
Infomax 2.1/500; Ext-Infomax 3.2/500 \newline
\textbf{Time:} 35.4\,min &
Component maps; runtime logs; status \\

\addlinespace[2pt]

Task~4: TUSL EEG processing &
300 files; 0.5--40\,Hz band-pass; SR 250\,Hz; windows 2.0\,s (0.5 overlap) &
\textbf{Validity:} 296 / 4 (valid / rejected); chans 21--23 \newline
\textbf{Confusion:} TP/FP/FN/TN = 14/5/86/191 \newline
\textbf{Time:} 15.6\,min &
\textbf{Validity:} 298 / 2 (valid / rejected); chans 21--23 \newline
\textbf{Confusion:} TP/FP/FN/TN = 20/11/62/203 \newline
\textbf{Time:} 51.9\,min &
Manifests; preprocessing logs; features \\

\bottomrule
\end{tabularx}
\end{table}

From a professional assessment perspective, these results highlight that the agent enforces \emph{quality-aware execution} rather than only automating scripts.
It couples quantitative checks (spectral SNR, variance explained, runtimes/iterations, yield/consistency statistics, and confusion counts) with interpretable diagnostics (topographic patterns, PSD, pre/post evoked plots, ICA component maps, and per-file audit logs), reducing the chance of silently accepting spurious or numerically unstable outputs.
We also document edge cases and limitations to strengthen auditability: if oscillatory structure is weak, spectral ratios may flatten and extracted patterns become diffuse, triggering warnings and preserving intermediate artifacts; missing optional resources (e.g., Maxwell filtering files or optional ICA backends) are handled by best-effort execution with structured logs; and for large-scale clinical EEG, channel variability and class imbalance can shift feature distributions, motivating subject/session-level calibration or ROC/PR-based threshold analysis, which the agent supports by exporting subject-level feature statistics and prediction traces.

\subsubsection{Summary}
Beyond these evaluated settings, the same agent design supports a wide range of electrophysiology scenarios without changing the overall workflow structure.
This is possible because the pipeline is modular: standardized ingestion, parameter normalization, preprocessing, logging, and delivery are reused, while the analysis and validation stages can be swapped or extended according to the research question.
For example, the system can be adapted to event-related potential studies by changing event alignment rules, epoch windows, and component extraction targets, while reusing the same data harmonization and reporting backbone.
It can support brain network modeling by enabling connectivity estimation and network metrics over chosen parcellations.
It can be extended to sleep analysis by adding oscillation-based staging heuristics and cycle detection, and to brain--computer interface workflows by integrating feature optimization and classifier training modules.
Clinical EEG use cases such as epileptiform activity screening can be handled by enabling anomaly detection components with domain-specific thresholds and reporting templates.
Importantly, these expansions do not require rewriting pipelines from scratch: the agent reuses the same ingestion, preprocessing, audit trails, and deliverable packaging mechanisms, and only changes the task-specific analysis logic and validation rules.
Figure~\ref{fig:eeg_case} illustrates this end-to-end behavior, summarizing how the agent translates a user request into a normalized configuration, explores and harmonizes heterogeneous datasets, executes a standardized preprocessing backbone, routes into scenario-specific analysis modules, applies multi-view validation, and exports a complete package of processed data and scientific reports.

\subsection{Case 3: Cross-Disciplinary Earth Data Preparation}
\label{subsec:polar_case}
\begin{figure}[t]
    \centering
    \includegraphics[width=1.0\textwidth]{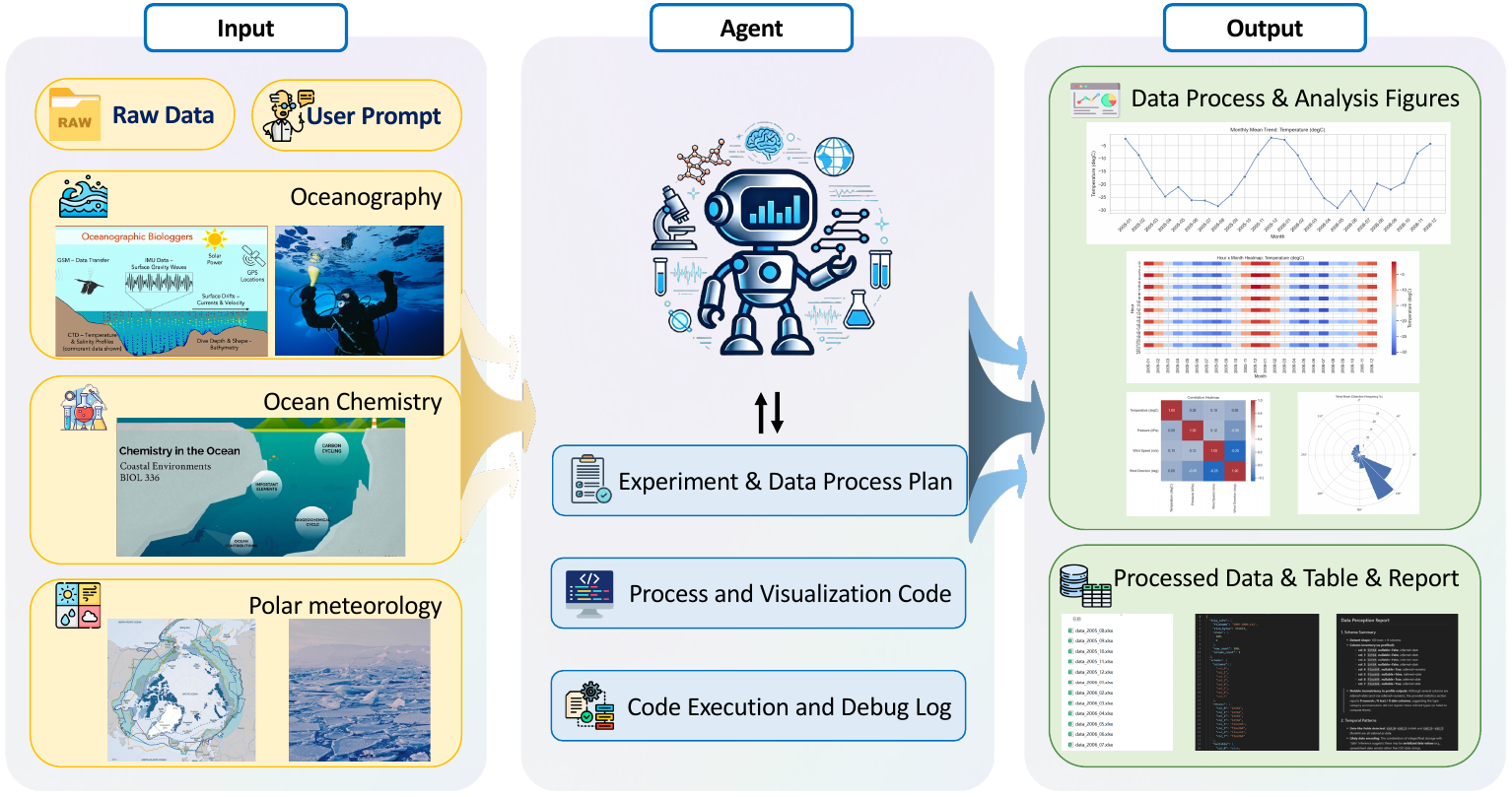}
    \caption{Case Study Diagram of the SciDataCopilot on Oceanography, Ocean Chemistry, and Polar Meteorology. }
    \label{fig:polar_case}
\end{figure}
\subsubsection{Problem Statement}
Long-term earth scientific analysis increasingly depends on the integration of large-scale, heterogeneous observations spanning physical oceanography, marine chemistry, sedimentology, cryosphere processes, and atmospheric environments into unified, analysis-ready datasets. However, traditional data curation remains constrained by disciplinary boundaries and format differences, often requiring cross-group collaboration among multiple departments and domain specialists. In practice, data cleaning and standardization for a single year can require weeks of sustained manual effort.

Historical earth archives are further characterized by systematic misalignment across multiple dimensions, including data format, temporal resolution, and semantic definition. These inconsistencies prevent heterogeneous observational sources from being jointly analyzed or directly consumed by data-driven and physics-informed AI models.

A central obstacle is that many legacy observations are not organized for downstream integration. Observational values are frequently stored in weakly structured files, while semantic headers and temporal fields are encoded separately or recorded in inconsistent conventions. As a result, constructing a coherent table with correct column meaning and standardized time indexing often requires repeated manual inspection and informal transformations.

From a data-centric AI perspective, this alignment step is a prerequisite: earth observations must first be converted into a semantically coherent and temporally standardized representation before cross-source and cross-modal integration can be performed. Without this transformation, it remains difficult to build reliable Scientific AI-Ready datasets that support reproducible analysis and modeling.

\subsubsection{Experimental Setting}
In this case study, we evaluate a representative alignment task in which observational values and semantic headers are stored in separate legacy files, and temporal organization does not match downstream analytical requirements. Concretely, the evaluation uses meteorological observations collected at Marble Point during 2005-2006, comprising 5840 rows\cite{wang2023antaws}, with a user instruction of "Process polar tabular data: merge header and records, compute daily averages from hourly values, then split outputs by month".

This task is selected because it isolates a foundational capability that is required for broader earth data integration. It evaluates whether a single natural-language specification can be translated into a deterministic, auditable processing pipeline that reconstructs column semantics, normalizes time fields, applies temporal aggregation, and exports standardized outputs that can serve as an upstream input to multi-source and multi-modal alignment.

\subsubsection{Experimental Results}
We conducted a comparative study to evaluate the efficiency and effectiveness of SciDataCopilot on Marble Point dataset.
The task involved processing 5840 rows of meteorological data to strictly follow the temporal partitioning and averaging requirements.
The agent successfully interpreted the natural language instruction and executed the complete data processing pipeline in just 3.5\,min.
In contrast, a proficient human operator, utilizing Excel with autocomplete functions, required 75\,min to complete the identical task.
These results demonstrate that the intelligent agent achieves an efficiency improvement of over $20\times$ compared to the traditional manual workflow.
Crucially, the agent not only accelerates the process but also ensures the generation of high-quality, standardized outputs that are immediately ready for downstream analysis, effectively eliminating the variability and potential errors associated with manual data handling.

\subsubsection{Summary}
The presented workflow follows the agentic data preparation design of SciDataCopilot, which converts fragmented raw files into Scientific AI-Ready data through coordinated agent roles. An intent parsing component translates the user request into an executable plan with explicit operations and constraints, while a data access component identifies relevant files, extracts observational matrices and metadata artifacts, and normalizes them into a canonical internal representation. A data processing component then performs semantic reconstruction and temporal reorganization under iterative checking and execution, ensuring that intermediate products remain traceable and that failures are handled through structured diagnostics rather than silent corruption.

Although the evaluation uses a meteorological table, the same interaction pattern extends to heterogeneous earth observations that differ in modality and storage conventions. The same pipeline-level structure can ingest and normalize cruise-based measurements, moored instrument records, and sediment sampling outputs, while reusing consistent mechanisms for schema reconstruction, temporal standardization, logging, and deliverable packaging. In this sense, the case should be viewed as a prerequisite module for multi-disciplinary integration, enabling subsequent steps such as cross-source variable matching and multi-modal fusion across ocean, ice, atmosphere, and biology.

\section{Conclusion}
\label{sec:conclusion}
In summary, we introduced \textbf{SciDataCopilot}, a unified agentic framework that bridges the gap between heterogeneous experimental data and the rigorous requirements of AI-driven scientific discovery.
By advancing the data preparation paradigm from AI-Ready to Scientific AI-Ready, the system explicitly addresses the challenges of task dependency, data fragmentation, and domain-specific processing constraints. The framework orchestrates four collaborative agents to automate the entire lifecycle of data preparation, transforming raw, messy data into Scientific AI-Ready resources.

Extensive deployments across diverse scientific domains confirm the system's effectiveness and generalization.
Across three use cases in life science, neuroscience, and earth science, SciDataCopilot validates the Scientific AI-Ready data paradigm by reliably transforming heterogeneous raw data into standardized, task-aligned resources under realistic scientific constraints.

Looking forward, several directions remain open. First, we plan to extend SciDataCopilot to a broader range of scientific domains and data modalities, including large-scale simulation outputs and real-time experimental streams, to further stress-test the generality of the Scientific AI-Ready paradigm. 
Second, while the current system relies on curated domain tools and cases, future work will explore more adaptive mechanisms for tool discovery, validation, and learning from execution traces, enabling continual improvement with minimal human intervention. 
Finally, we aim to formalize evaluation protocols for Scientific AI-Ready data beyond efficiency and correctness, incorporating measures of scientific validity, reproducibility, and downstream impact on scientific modeling and discovery.



\begingroup
\sloppy
\clearpage
\printbibliography[heading=bibintoc]
\endgroup


\end{document}